\documentclass[aps,amsmath,twocolumn,amssymb,floatfix,showpacs,superscriptaddress,nofootinbib,longbibliography]{revtex4-1}
\usepackage{mathtools}
\usepackage{braket}
\usepackage[dvipsnames]{xcolor}
\usepackage{float}
\usepackage{mathdots}
\usepackage{subfigure}
\usepackage[colorlinks=true,linktoc=page,linkcolor=Purple,citecolor=blue,urlcolor=RedViolet]{hyperref}

\newcommand{\vect}[1]{\boldsymbol{\mathrm{#1}}}
\mathchardef\mhyphen="2D 

\newcommand{\ie}{{i.e.,\,\,}}
\newcommand{\eg}{{e.g.,~}}
\newcommand{\ua}{{\uparrow }}
\newcommand{\da}{{\downarrow }}

\newcommand{\la}{{\langle}}
\newcommand{\ra}{{\rangle}}

\newcommand\bea{\begin{eqnarray}}
\newcommand\eea{\end{eqnarray}}
\newcommand\beq{\begin{equation}}  
\newcommand\eeq{\end{equation}}

\newcommand{\non}{\nonumber}  
\usepackage[normalem]{ulem}
\definecolor{lime}{HTML}{A6CE39}
\usepackage{sidecap,tikz}
\DeclareRobustCommand{\orcidicon}{\hspace{-1.0mm}
	\begin{tikzpicture}
		\draw[lime, fill=lime] (0.0,0.0) 
		circle [radius=0.15] 
		node[white] {{\fontfamily{qag}\selectfont \tiny \,ID}};
		\draw[white, fill=white] (-0.0525,0.095) 
		circle [radius=0.007];
	\end{tikzpicture}
	\hspace{-3.0mm}
}
\foreach \x in {A, ..., Z}{\expandafter\xdef\csname orcid\x\endcsname{\noexpand\href{https://orcid.org/\csname orcidauthor\x\endcsname}
		{\noexpand\orcidicon}}
}

\AtBeginDocument{%
	\newwrite\bibnotes
	\def\bibnotesext{Notes.bib}
	\immediate\openout\bibnotes=\jobname\bibnotesext
	\immediate\write\bibnotes{@CONTROL{REVTEX41Control}}
	\immediate\write\bibnotes{@CONTROL{%
			apsrev41Control,author="08",editor="1",pages="1",title="1",year="1"}}
	\if@filesw
	\immediate\write\@auxout{\string\citation{apsrev41Control}}%
	\fi
}%
\begin{document}


\title{Transport signatures of single and multiple Floquet Majorana modes in one-dimensional Rashba nanowire and Shiba chain}

\author{Debashish Mondal\orcidD{}}
\email{debashish.m@iopb.res.in}
\affiliation{Institute of Physics, Sachivalaya Marg, Bhubaneswar-751005, India}
\affiliation{Homi Bhabha National Institute, Training School Complex, Anushakti Nagar, Mumbai 400094, India}

\author{Rekha Kumari}
\email{rekha.kumari@icts.res.in}
\affiliation{International Centre for Theoretical Sciences, Tata Institute of Fundamental Research, Bengaluru 560 089, India}

\author{Tanay Nag\orcidB{}}
\email{tanay.nag@hyderabad.bits-pilani.ac.in}
\affiliation{Department of Physics, BITS Pilani-Hyderabad Campus, Telangana, 500078, India}

\author{Arijit Saha\orcidC{}}
\email{arijit@iopb.res.in}
\affiliation{Institute of Physics, Sachivalaya Marg, Bhubaneswar-751005, India}
\affiliation{Homi Bhabha National Institute, Training School Complex, Anushakti Nagar, Mumbai 400094, India}

\begin{abstract}
We theoretically investigate the transport signature of single and multiple Floquet Majorana end modes~(FMEMs), appearing in an experimentally feasible setup with Rashba nanowire~(NW) placed in closed proximity to a conventional $s$-wave superconductor, in the presence of an external Zeeman field. Periodic drive causes the anomalous $\pi$-modes to emerge in addition to the regular $0$-modes in the driven system where the former does not exhibit any static analog. For single $0$- and/or $\pi$-FMEM, differential conductance exhibits a quantized value of $2e^{2}/h$ while we consider the sum over all the photon sectors, supporting Floquet sum rule. We examine the stability of this summed conductance against random onsite disorder. We further investigate the summed conductance in several cases hosting multiple~(more than one) $0$- or $\pi$-modes at the end of the NW. In these cases, we obtain quantized values of $n_{M}\times 2e^{2}/h$ in summed differential conductance with $n_{M}$ being the number of modes~($0$ or $\pi$) localized at one end of the NW. We repeat our analysis for another experimentally realizable model system known as helical Shiba chain. Moreover, we corroborate our results by computing the differential conductance for FMEMs 
using non-equilibrium Green's function method. Our work opens up the possibility of studying the transport signatures of FMEMs in these realistic models.
\end{abstract}

\maketitle

\section{Introduction}
In recent times, Majorana fermions~(MFs) in the form of Majorana zero modes~(MZMs) associated with topological superconductor~\cite{Kitaev_2001,qi2011topological,Leijnse_2012,Alicea_2012,ramonaquado2017,beenakker2013search,Flensberg2021,10.1093/ptep/ptae065} have received enormous attention due to their non-Abelian braiding properties. This suffices them 
to serve as the fundamental unit for topological quantum computations~\cite{Ivanov2001,freedman2003topological,KITAEV20032,Stern2010,NayakRMP2008}. Initially, Kitaev introduced the concept of MZMs in his seminal theoretical paper, proposing their existance at the ends of a one-dimensional (1D) spinless $p$-wave superconducting chain~\cite{Kitaev_2001}. However, the inherent challenge of accessing $p$-wave superconductivity has hindered the experimental realization of Kitaev's proposal. Nevertheless, an experimentally viable configuration involving a 1D Rashba nanowire (NW) in close proximity to a regular $s$-wave superconductor under the influence of an external Zeeman field could effectively emulate a 1D $p$-wave superconductor~\cite{Oreg2010,LutchynPRL2010,Leijnse_2012,Alicea_2012,Mourik2012Science,das2012zero,ramonaquado2017}. Based on semiconducting NW-superconductor heterostructure, 
zero bias peak in differential conductance has been observed in various transport experiments which manifest the indirect signatures of the MZMs~\cite{Mourik2012Science,das2012zero,Rokhinson2012,Finck2013,Albrecht2016,Deng2016}. In this direction, there exists an alternative approach based on helical spin chain~\cite{Yazdani2013,Felix_analytics,Loss2013PRL,Felix2014,PhysRevB.89.115109, Bena2015,Loss2016,Christensen2016,Simon2017,Sticklet2019,Rex2020,Mohanta2021, Loss2022,PritamPRB2023} 
in which magnetic adatoms are implanted on the surface of a bulk $s$-wave superconductor
~\cite{Sau2013,PhysRevB.88.180503,Hui2015,Joel2015,Sharma2016,Theiler2019,Mashkoori2020,Teixeria2020,Bedow2022}. Here, MZMs appear within the minigap of emergent Shiba bands. The existence of MZMs 
in such setup has been experimentally realized via several recent experiments~\cite{Yazdani_science,Yazdani2015,HowonSciAdv2018,Schneider2020,Wiesendanger2021,Beck2021,Wang2021PRL,Schneider2022,Crawford2022} using scanning tunneling microscopy~(STM) measurements. 

Having discussed the static/equilibrium signatures, we now focus on the theoretical understanding of a driven topological superconductor where experimental progress is still in it's infancy. 
In current literature, Floquet engineering represents an efficient and advanced route for tailoring desired topology in a non-topological system~\cite{oka09photovoltaic,kitagawa11transport,lindner11floquet,Rudner2013,Usaj2014,Piskunow2014,Eckardt2017,Yan2017,oka2019,NHLindner2020,nag2021anomalous,ThakurathiPRB2013,benito14,PotterPRX2016,JiangPRL2011,ReynosoPRB2013,LiuPRL2013,ThakurathiPRB2017,MitraPRB2019,YangPRL2021,sacramento15,rxzhang21,PhysRevB.100.041103,PhysRevB.101.155417,An_hybrid_2023, Li_PRB_2014, Wenk_PRB_2022, Force_PRB_2023, BENITO2015608}. The intricate winding of the time-dependent wave function results in the inception of anomalous topological boundary modes at finite quasi-energy, termed as $\pi$-modes leaving no static equivalence. One can have control over the number of emerging multiple Floquet Majorana end modes (FMEMs) by modulating the amplitude and frequency of the drive. Emergence of multiple MZMs has been explored in $p$-wave Kitaev chain~\cite{ThakurathiPRB2013, benito14, PotterPRX2016, ton_PRB_2013, Wu_NJP_2023, Ahmed_2024}, 
1D cold-atomic NW-$s$-wave superconductor heterostructure~\cite{JiangPRL2011,ThakurathiPRB2017,LiuPRL2013,YangPRL2021,MitraPRB2019}, realistic 1D Rashba NW-superconductor hybrid setup~\cite{Mondal2023_NW} and also in helical Shiba chain model~\cite{Mondal_2023_Shiba}. The possibility of braiding operation with these out of equilibrium Floquet modes further gives rise to significant future research avenues in the direction of topological quantum computations~\cite{Bomantara18,BomantaraPRB2018,BelaBauerPRB2019,MatthiesPRL2022}.

In this direction, the transport signature of FMEMs was first explored in 1D Kitaev model by executing Floquet sum rule to obtain quantized $2e^{2}/h$ peak for summed conductance $\tilde{\sigma}$~\cite{KunduPRL2013}. The robustness of the quantized peaks against disorder was also emphasized. In the presence of periodic driving, the concept of the Floquet sum rule is inspected to other setups such as quantum well heterosturcture~(a two-dimensional (2D) quantum spin Hall insulator)~\cite{Pereg_PRL_2015} and topological insulator~\cite{Pereg_PRB_2016} leading to non-quantized value~($<2e^{2}/h$ or $>2e^{2}/h$) of summed differential conductance $\tilde{\sigma}$ at zero bias. In this case, it has been shown that the zero bias peak (ZBP) $<2e^{2}/h$ values are robust against disorder while others $>2e^{2}/h$ are not. Further investigation in this direction leads to a planner Josephson junction with proximitized heterosturcture based on 2D electron gas with Rashba spin-orbit coupling~(SOC) and Zeeman field~\cite{MitraPRB2019}. Floquet transport is also studied in other sytems as well~\cite{Yap_PRB_2018, Simons_PRB_2021}.  However, the transport signatures of FMEMs, appearing in experimentally realizable Rashba NW model and helical Shiba chain model, are yet to be explored to the best of our knowledge. Detection of FMEMs in these systems is crucial for any practical application, and transport serves as an effective tool for this purpose. This motivates us to examine the transport signatures of dynamical Majorana modes once their number can be tuned. At first, we generalize the theory of Floquet transport for FMEMs~\cite{KunduPRL2013} consideirng the systems possessing spin/chiral degrees of freedom~(DOF). Then the intriguing questions that we address here are the following: How do Floquet sidebands contribute to the quantized signal of differential conductance in such Rashba NW-parent $s$-wave superconductor hybrid setup? How can we understand the relation between the stability and quantization of regular and anomalous Majorana modes in the presence of disorder? What is the role of the bulk gap associated with FMEMs as far as the stability of quantized differential conductance against disorder is concerned? How does the differential conductance behave when multiple FMEMs are localized at one end of the concerned system? 

\begin{figure}[]
	\centering
	\subfigure{\includegraphics[width=0.5\textwidth]{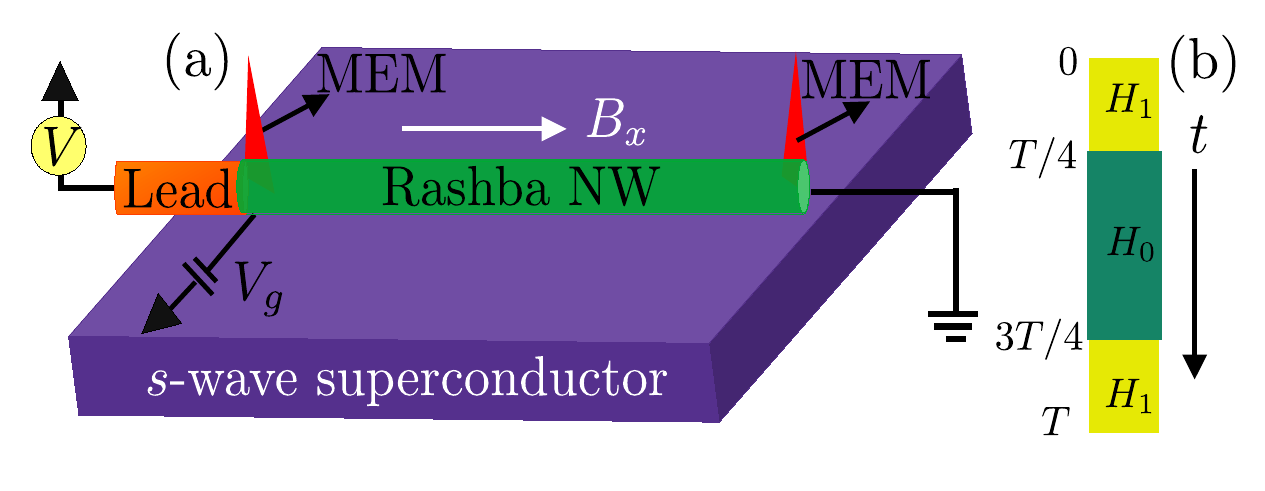}}
	\caption{(a) Schematic diagram of our setup is presented to study transport signature of FMEMs in Rashba nanowire~(NW) model. Here, a 1D NW~(green) with strong Rashba SOC is placed on top of a common $s$-wave superconducting substrate~(violet) in the presence of an external magnetic field $B_x$ applied along the length of the NW~($x$-direction). A pair of MZMs~(red) emerges at the ends of the NW. To tune the chemical potential a gate voltage $V_{g}$ is applied to the NW. A metallic lead~(orange) is attached to one end of the NW and a bias voltage $V$ is applied across the lead to measure single terminal differential conductance while another end of the NW is connected to the ground. (b) Three-step drive protocol, according to Eq.~(\ref{eq:step_drive}), 
is schematically depicted here to generate FMEMs.}
	\label{fig:schematic_NW}
\end{figure}

In this article, first, we begin with the experimentally feasible 1D Rashba NW model in close proximity of a bulk $s$-wave superconductor~(see Fig.~\ref{fig:schematic_NW} for our schematic setup). 
The heterostructure manifests itself as an effective $p$-wave superconductor by hosting a pair of MZMs at the ends of the NW in the topological regime. For completeness, we study the transport 
property of these static zero-energy MEMs. We demonstrate the differential conductance $\sigma_{\rm stat}$ as a function of bias voltage $V$ where we obtain the quantized ZBP value of $2e^{2}/h$ 
for $\sigma_{\rm stat}$ at $V=0$ (see  Fig.~\ref{fig:static_NW}) as an indirect signature of MZMs. To investigate the transport signature of FMEMs, we apply three-step periodic drive protocol that results 
in regular $0$- and anomalous $\pi$-FMEMs in the system~\cite{Mondal2023_NW}. We calculate one-terminal differential conductance $\sigma$ for individual photon sectors, their sum $\tilde{\sigma}$ and represent them as a function of bias voltage $V$ for four cases where  single $0$- and/or $\pi$-modes are located at one end of the NW. For all these cases, we encounter quantized values of 
$\tilde{\sigma}$ corresponding to $0$- and $\pi$-modes. However, the individual photon sector does not yield quantized response unless the Floquet sum rule over all the photon sectors is satisfied~ \cite{KunduPRL2013} ~(see  Fig.~\ref{fig:photon_NW}). We check the robustness of $\tilde{\sigma}$ against random onsite disorder for all the above instances,~(see  Fig.~\ref{fig:disorder}). We also explore the cases when multiple~(more than one) $0$- and/or $\pi$- modes are present at one end of the NW~(see Fig.~\ref{fig:multiple}). We extend our investigation to another practically realizable model based on magnetic adatoms on the surface of an $s$-wave superconductor~namely, helical Shiba chain model, and obtain qualitatively identical results akin to the previous model~(see Fig.~\ref{fig:schematic_Shiba} for schematic representation and Fig.~\ref{fig:Shiba_result} for the results). After a complete study of the transport signature of FMEMs for both the realistic models, we adopt another numerical approach based on non-equilibrium Green's function~(NEGF) method to validate our results further (see Fig.~\ref{fig:NW_NEGF} and Fig.~\ref{fig:Shiba_NEGF}). We also provide possible experimental feasibility of our theoretical findings.

The remaining parts of our article are arranged as follows. We introduce the static Hamiltonian for Rashba NW model in Sec.~\ref{sub:mod_NW} and demonstrate the  driving protocol to engineer FMEMs  in Sec~\ref{sub:step}. In Sec~\ref{sec:fl_th}, we briefly review the Floquet theory needed to examine the differential conductance. We briefly discuss the transport theory for Floquet and static MEMs in Secs.~\ref{sec:tr_fl_th} and ~\ref{sec:tr_st_th}, respectively. We discuss the transport results for static Rashba NW in Sec.~\ref{sub:stat_result} for completeness. In case of driven Rashba NW, we illustrate our results for single FMEM in Sec.~\ref{sub:single_NW} and their stability against disorder in Sec.~\ref{sub:dis}. In Sec.~\ref{sub:multi}, we present the same for multiple FMEMs. Sec.~\ref{sec:Shiba} is 
devoted to the transport study of static and Floquet MEMs in case of helical Shiba chain model. We present our results using NEGF technique in Sec.~\ref{sec:NEGF_result}. The choice of parameters for 
our numerical findings and the corresponding experimental feasibility of our setups are described in Sec.~\ref{sec:experiment}. In Sec.~\ref{sec:discussion}, we add a discussion on our transport results and possible future directions. Finally, we summarise and conclude our article in Sec.~\ref{sec:conclusion}.

\section{Model and driving protocol}\label{Sec:II}
\subsection{Model Hamiltonian}\label{sub:mod_NW}
We consider a 1D Rashba NW placed on the top of an $s$-wave superconductor in presence of a Zeeman field applied parallel to the NW~\cite{Alicea_2012,Leijnse_2012,LutchynPRL2010,Oreg2010,Mondal2023_NW}. Superconductivity is induced in the NW as a consequence of the proximity effect. Let us consider the Bogoliubov-de Gennes (BdG) basis: $\Psi_{j}=\{\psi_{j\ua},\psi_{j\da},\psi_{j\da}^{\dagger},-\psi_{j\ua}^{\dagger}\}^{\mathbf{t}}$ with $\psi_{j\ua}^{\dagger}$~($\psi_{j\ua}$) and $\psi_{j\da}^{\dagger}$~($\psi_{j\da}$) stand for the creation~(anihilation) operator at j\textsuperscript{th} site for spin-up and spin-down sectors, respectively, while $\mathbf{t}$ represents the transpose operation. Then BdG Hamiltonian for this setup can be written as~\cite{Mondal2023_NW}
\begin{eqnarray}
	H_{0}&=&\sum_{j=1}^{N} \Psi_{j}^{\dagger}\left[ (2t_{h}-c_{0})\Gamma_{1} + B_{x} \Gamma_{3} +\Delta \Gamma_{4} \right] \Psi_{j} \nonumber  \\
	&+& \sum_{j=1}^{N-1} \Psi_{j+1}^{\dagger} \left[-t_{h} \Gamma_{1} -iu \Gamma_{2}\right] \Psi_{j} + H.c.\ , \label{eq:Hamiltonian}
\end{eqnarray}
where, the $4\times4$-${\Gamma}$-matrices are given by $\Gamma_{1}=\tau_{z}\sigma_{0}$, $\Gamma_{2}=\tau_{z}\sigma_{z}$, $\Gamma_{3}=\tau_{0}\sigma_{x}$, and $\Gamma_{4}=\tau_{x}\sigma_{0}$ with the Pauli matrices $\vect{\tau}$ and $\vect{\sigma}$ act on particle-hole and spin subspaces, respectively. Here $c_{0}$, $t_{h}$, $B_{x}$, $u$, and $\Delta$ denote chemical potential, nearest neighbour hopping amplitude, magnetic field, Rashba SOC strength and proximity induced $s$-wave superconducting gap inside the NW, respectively. This system exhibits topological phase boundaries given by $\sqrt{c_{0}^{2}+\Delta^{2}}\leq |B_{x}| \leq \sqrt{(4t_{h}-c_{0})^{2}+\Delta^{2}}$~\cite{Mondal2023_NW}. In the topological regime, under open boundary condition~(OBC), the NW hosts one Majorana $0$-mode at each of its ends~\cite{Mondal2023_NW}.

\subsection{Driving protocol}\label{sub:step}
In order to Floquet band engineering of our system, we consider the following 3-step periodic drive protocol reads as
\begin{eqnarray}
	H(t)&=&H_{1}=\sum_{j=1}^{N} \Psi_{j}^{\dagger} \left[-c_{1} \Gamma_{1} \right] \Psi_{j} \hspace*{0.3 cm} t\in \left[0,\frac{T}{4}\right)\ , \nonumber \\ 
	&=& H_{0} \hspace*{3.5 cm} t\in \left[\frac{T}{4}, \frac{3T}{4}\right)\ , \nonumber	\\
	&=&H_{1}= \sum_{j=1}^{N} \Psi_{j}^{\dagger} \left[-c_{1} \Gamma_{1} \right] \Psi_{j} \hspace*{0.3 cm} t\in \left[\frac{3T}{4},T\right] \label{eq:step_drive},
\end{eqnarray} 
where, $H_{1}$ term denotes the modulation of the onsite chemical potential only and $T$~(=$2\pi/\Omega$) is the time period of the drive. Starting from the non-topological regime of the static Hamiltonian~[Eq.~(\ref{eq:Hamiltonian})] we can generate regular Floquet $0$- and anomalous $\pi$-modes appearing in our system due to the application of the above mentioned drive~[Eq.~(\ref{eq:step_drive})]. By changing the frequency $\Omega$ and amplitude $c_{1}$ of the drive, one can also achieve control over the number of these emergent FMEMs~\cite{Mondal2023_NW}.

\section{Floquet transport Theory}\label{sec:theory}
In this section, we briefly discuss the Floquet theory for investigating the transport properties of FMEMs and static MZMs. However, the details are discussed in Appendix~\ref{sec:notations}, Appendix~\ref{sec:primer_Floquet} and Appendix~\ref{sec:app_transport}.

\subsection{Floquet theory}\label{sec:fl_th}
For a periodically driven system \ie $H(t+T)=H(t)$, following the analogy with Bloch theorem, the solution of the Schr\"{o}dinger's equation $H(t) | \psi_{\bar{\alpha}}(t) \rangle= i \partial_{t} | \psi_{\bar{\alpha}} (t)\rangle$ takes the form $| \psi_{\bar{\alpha}} (t)\rangle= e^{-i \epsilon_{\bar{\alpha}}t} | u_{\bar{\alpha}} (t)\rangle$ with $| u_{\bar{\alpha}} (t)\rangle=| u_{\bar{\alpha}} (t+T)\rangle$. Here, $| u_{\bar{\alpha}} (t)\rangle$ coined as Floquet states are the eigenvectors of effective Hamiltonian $H_{\rm{eff}}=H(t)-i \partial_{t}$, $H_{\rm{eff}}| u_{\bar{\alpha}} (t)\rangle= \epsilon_{\bar{\alpha}} | u_{\bar{\alpha}} (t)\rangle$  and corresponding eigenvalues $\epsilon_{\bar{\alpha}}$ are called the quasi-energies~\cite{Eckardt_2015,Rodriguez-Vega_2018}. With the help of the periodicity of 
$|\psi(t) \rangle$, the time evolution operator $U(T,0)$ satisfies the following relation: $U(T,0)  \ket{u_{\bar{\alpha}}(T)}  = e ^{-i \epsilon_{\bar{\alpha}} T} \ket{u_{\bar{\alpha}}(T)}$. From the eigenvalues 
$\Lambda_{\bar{\alpha}}$ of $U(T,0)$, we compute quasi-energies $\epsilon_{\bar{\alpha}}=(i/T)\ln(\Lambda_{\bar{\alpha}})$.  $ \epsilon_{\bar{\alpha}} \rightarrow \epsilon_{\bar{\alpha}} + n \Omega$ with $n \in \mathbb{Z}$ provides a new set of $\{\ket{u_{\bar{\alpha}}(t)}\} \hspace*{ 1mm} \rightarrow \hspace*{1mm} \{ e^{i n \Omega t} \ket{u_{\bar{\alpha}}(t)}\}$ for the same set of $\{\ket{\psi_{\bar{\alpha}}(t)}\}$. Consequently, the quasi-energies are not unique and are connected via the periodic drive by the absorption and emission of virtual photons. As a result, these frequency indices are commonly referred to as photon indices (side bands) within the context of Floquet theory~\cite{Eckardt_2015,KunduPRL2013}. Nonetheless, quasi-energies maintain uniqueness within the realm of the first Floquet zone or the $0$\textsuperscript{th} photon sector:$-\Omega/2 \leq \epsilon_{\bar{\alpha}} \leq \Omega/2$. Fourier decomposition of the periodic $\ket{u_{\bar{\alpha}}(t)}$ is given by $\ket{u_{\bar{\alpha}}(t)} = \sum_{n} e^{-i n \Omega t} \ket{u_{\bar{\alpha}}^{(n)}}$ and related frequency space Schr\"{o}dinger's equation takes the form: $\left[ H^{(k-n)}-k\Omega \right]|u_{\bar{\alpha}}^{(n)}\rangle = \epsilon_{\bar{\alpha}} |u_{\bar{\alpha}}^{(k)}\rangle$ with $H^{(k-n)}= \int_{0}^{T}\frac{dt}{T} e^{i (k-n)\Omega t} H(t) $. Note that, $|u_{\bar{\alpha}}^{(k)}\rangle$ constitute a Hilbert space called 
extended Hilbert space (see Appendix~\ref{sec:Extended space Hamiltonan}) where expectation value of an observable ${\cal{O}}$ is defined as $\la \la {\cal{O}}\ra \ra=\sum_{k}\la u_{\bar{\alpha}}^{(k)}|{\cal{O}}|u_{\bar{\alpha}}^{(k)}\rangle$~\cite{KunduPRL2013}. See Appendix~\ref{sec:primer_Floquet} for more details.
\subsection{Transport theory of Floquet Majorana modes}\label{sec:tr_fl_th}
For transport study of FMEMs, we consider the setup as mentioned in Fig.~\ref{fig:schematic_NW}. One end of the system~(Rashba NW) is connected to a single channel metallic lead attached to a thermal reservoir. We apply bias voltage via this lead and another end of the NW is connected to the ground. Here, we assume that the lead follows the Fermi-Dirac distribution function and does not change with time indicating the lead density of states $\rho$ to be constant. The attached lead introduces an imaginary self-energy $\delta_{\bar{\alpha}}$ to the quasi-energy $\epsilon_{\bar{\alpha}}$ 
of the driven system. For bias voltage $V$, one terminal~(single lead) differential conductance is given by~(see Appendix~[\ref{sec:app_transport}] for detailed discussion)
\begin{eqnarray}
	\sigma(V) &=&  -2 \pi e^{2} \!\! \int \! d\omega \! \sum_{k} \! \operatorname{Tr}\left[\mathbf{G}^{(k)\dagger}(\omega)  \mathbf{V}^{\mathbf{t}}\mathbf{G}^{(k)}(\omega) \mathbf{V} \right]\  \nonumber \\
	&&\times\left(f^{'}(\omega) +f^{'}(-\omega) \right) \label{eq:main_mu_1_BDG},
\end{eqnarray}
where $f$ represents the Fermi-Dirac distribution function of the reservoir, and $G^{(k)}$ is the Nambu-Gorkov Green's function given by~\cite{MitraPRB2019}
\begin{eqnarray}
	\mathbf{G}^{(k)}(\omega)&=& \sum_{\bar{\alpha} ,n} \frac{|u_{\bar{\alpha}}^{(k+n)}\rangle \langle u_{\bar{\alpha}}^{(n)}|}{\omega-\epsilon_{\bar{\alpha}}-n\Omega +i\delta_{\bar{\alpha}}} \label{eq:main_NEGF}\ .
\end{eqnarray}
Here, $\delta_{\bar{\alpha}}$ is self energy correction to quasi-energy $\epsilon_{\bar{\alpha}}$ due to interaction between the system and the lead. 
We define the coupling parameter $\nu= 2 \pi \rho t_{h}^{2}$. 
The $\mathbf{V}$ matrix, carrying the information of surface coupling between the lead and system,  vanishes everywhere except for the contact site \ie $\mathbf{V}=\mathbf{V}^{c} \oplus 0 \oplus 0 \oplus 0 ... $, with $\textbf{V}^{c}$ is given by~(in BdG basis)
\begin{equation}
	\mathbf{V}^{c}=\rho t_{h}^{2} ~ \mathbf{I_{4}} \label{eq:main_V_4by4}\ .
\end{equation}
Here, $\mathbf{V}$ is a $4N \times 4N$ diagonal matrix out of which all the remaining
$N-1$ blocks are null matrix except for the first block designated by  $\mathbf{V}^{c}$.
In this limit Eq.~(\ref{eq:main_mu_1_BDG}) reduces to (see Appendix~[\ref{sec:app_transport}] 
for details)
\begin{widetext}
\begin{eqnarray}
	\sigma(V)
	&=& - \frac{e^{2} \nu^{2}}{2 \pi} \times2 \times \!\! \sum_{k,s,s^{\prime}} \int\!\!d\omega \!\! \hspace*{2 mm}  \left[ |G^{c(k)}_{ee,ss^{\prime}}(\omega)|^{2}+ |G^{c(k)}_{eh,ss^{\prime}}(\omega)|^{2}+ |G^{c(k)}_{he,ss^{\prime}}(\omega)|^{2} + |G^{c(k)}_{hh,ss^{\prime}}(\omega)|^{2} \right]  f^{'}(\omega)\ , \! \label{eq:main_mu_2_BDG}
\end{eqnarray}
\end{widetext}
where, $G^{c(k)}_{he,ss^{\prime}}$ is the hole~(with $s$-spin/chirality)-electron~(with $s^{\prime}$-spin/chirality) component  of Nambu-Gorkov Green's function for photon sector $k$ computed 
at the contact site. Note that, in Eq.~(\ref{eq:main_mu_2_BDG}) we now have contribution arising
from both normal and Andreev reflection.
Moreover, weak coupling approximation~($\nu \ll t_{h}$) between the system and lead motivates one to perform perturbative analysis to obtain the self-energy as~\cite{KunduPRL2013}:
\begin{eqnarray}
	\delta_{\bar{\alpha}} &=& - \la \la -\pi \left(\mathbf{V}+\mathbf{V}^{\textbf{t}} \right) \ra \ra \nonumber 	 \\
	&=&\nu~ \sum_{k,s} \left[ |u_{\bar{\alpha},s}^{c(k)}|^{2}+ |v_{\bar{\alpha},s}^{c(k)}|^{2}\right]
	\label{eq:main_delta_alpha}\ ,
\end{eqnarray}
where $u_{\bar{\alpha},s}^{c(k)}$, and $v_{\bar{\alpha},s}^{c(k)}$ denote the particle and hole contribution of the wave function, respectively, for spin/chiral sector $s$ and photon sector $k$ 
at the contact site. Hence, in the zero temperature limit, the one terminal differential conductance in presence of a bias voltage $V$ is given by
\begin{widetext}
\begin{eqnarray}
	\tilde{\sigma}(V)&=& \operatorname{lim}_{V \rightarrow \epsilon_{\bar{\alpha}}} \frac{e^{2} \nu^{2}}{2 \pi}\times2  \sum_{k,s,s^{\prime}} \left[ |G^{c(k)}_{ee,ss^{\prime}}(V+n\Omega)|^{2}+|G^{c(k)}_{eh,ss^{\prime}}(V+n\Omega)|^{2}+|G^{c(k)}_{he,ss^{\prime}}(V+n\Omega)|^{2}+|G^{c(k)}_{hh,ss^{\prime}}(V+n\Omega)|^{2} \right] \nonumber \\
	&\approx& \!\!\!\sum_{n}\frac{2 e^{2}}{h} \!\!\! \sum_{\bar{\alpha},k,s,s^{\prime}} \! \frac{\nu^{2}}{\delta_{\bar{\alpha}}^{2}} \!\times \left[ |u_{\bar{\alpha},s}^{c (k+n)} u_{\bar{\alpha},s^{\prime}}^{c(n)}|^{2}+
	|u_{\bar{\alpha},s}^{c (k+n)} v_{\bar{\alpha},s^{\prime}}^{c(n)}|^{2}+|v_{\bar{\alpha},s}^{c (k+n)} u_{\bar{\alpha},s^{\prime}}^{c(n)}|^{2}+|v_{\bar{\alpha},s}^{c (k+n)} v_{\bar{\alpha},s^{\prime}}^{c(n)}|^{2} \right] \times \operatorname{L}\left( \frac{V-\epsilon_{\bar{\alpha}}}{\delta_{\bar{\alpha}}} \right) \non \\
	&=& \!\!\!\sum_{n} \sigma(V+n\Omega)\ , \label{eq:main_sum_slpit}
\end{eqnarray}
with contribution from $n\textsuperscript{th}$ photon sector $\sigma(V+n\Omega)=\frac{2 e^{2}}{h}  \sum_{\bar{\alpha},k,s,s^{\prime}} \frac{\nu^{2}}{\delta_{\bar{\alpha}}^{2}} \times \left[ |u_{\bar{\alpha},s}^{c (k+n)} u_{\bar{\alpha},s^{\prime}}^{c(n)}|^{2}+
|u_{\bar{\alpha},s}^{c (k+n)} v_{\bar{\alpha},s^{\prime}}^{c(n)}|^{2}+|v_{\bar{\alpha},s}^{c (k+n)} u_{\bar{\alpha},s^{\prime}}^{c(n)}|^{2}+|v_{\bar{\alpha},s}^{c (k+n)} v_{\bar{\alpha},s^{\prime}}^{c(n)}|^{2} \right] \times \operatorname{L}\left( \frac{V-\epsilon_{\bar{\alpha}}}{\delta_{\bar{\alpha}}} \right) \equiv \sigma^{(n)}(V)$. Here, the Lorentzian is defined by $\operatorname{L}(z)=\frac{1}{1+z^{2}}$.
\end{widetext}
\subsection{Transport through static Majorana modes in Rashba NW system} \label{sec:tr_st_th}
In order to calculate the transport signature of static MZMs, one can follow the above framework considering a static system. However, there exists a simple way to obtain the corresponding expression 
by considering the zero amplitude and zero frequency limit of the drive for the driven case. Thus, self-energy for the corresponding static system is given by

\begin{eqnarray}
	\delta_{\bar{\alpha}} &=&  \nu \sum_{s} \left[ |u_{\bar{\alpha},s}^{c}|^{2}+ |v_{\bar{\alpha},s}^{c}|^{2}\right]\ . \label{eq:delta_alpha_stat}
\end{eqnarray}
Then, the expression for one terminal differential conductance, caused by the static MZMs, is given by
\begin{widetext}
\begin{eqnarray}
	\sigma_{\rm{stat}}(V)\!\!&=&\!\frac{2 e^{2}}{h} \sum_{\bar{\alpha},s,s^{\prime}}  \frac{\nu^{2}}{\delta_{\bar{\alpha}}^{2}} \left[|u_{\bar{\alpha},s}^{c } u_{\bar{\alpha},s^{\prime}}^{c}|^{2}+ |u_{\bar{\alpha},s}^{c } v_{\bar{\alpha},s^{\prime}}^{c}|^{2}+  |v_{\bar{\alpha},s}^{c } u_{\bar{\alpha},s^{\prime}}^{c}|^{2}+|v_{\bar{\alpha},s}^{c } v_{\bar{\alpha},s^{\prime}}^{c}|^{2}\right] \times \! \operatorname{L}\left( \frac{V-E_{\bar{\alpha}}}{\delta_{\bar{\alpha}}} \right) \label{eq:main_stat_sigma}.
\end{eqnarray}
\end{widetext}

Here, $E_{\bar{\alpha}}$ denotes the $\bar{\alpha}$\textsuperscript{th} eigenvalue of the static Hamiltonian. To be precise, we obtain Eq.~(\ref{eq:delta_alpha_stat}) from Eq.~(\ref{eq:main_delta_alpha}) by excluding the frequency sum $k$. To obtain Eq.~(\ref{eq:main_stat_sigma}) from Eq.~(\ref{eq:main_sum_slpit}) we repeat the same and replace quasi-energy $\epsilon_{\bar{\alpha}}$ with static energy eigenvalues $E_{\bar{\alpha}}$.
\section{Results for Rashba NW setup}\label{sec:results}
\subsection{Transport signature of  static MZMs}\label{sub:stat_result}
\begin{figure}[]
	\centering
	\subfigure{\includegraphics[width=0.5\textwidth]{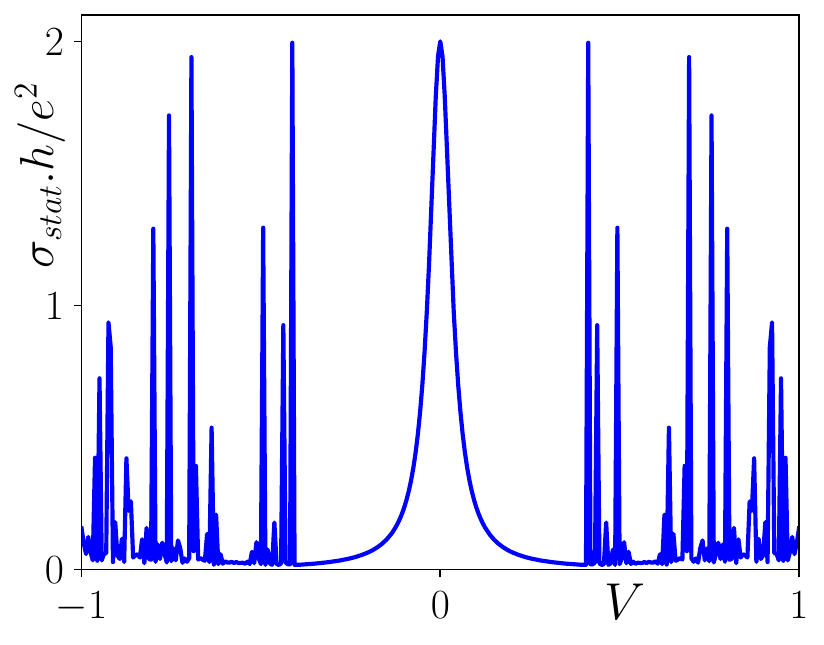}}
	\caption{We compute single-terminal differential conductance $\sigma_{\rm{stat}}$ using Eq.~(\ref{eq:main_stat_sigma}) for the static Rashba NW system in its topological regime. The zero bias~(V=0) peak of $\sigma_{\rm{stat}}$ exhibits a quantized value of $2 e^{2}/h$. We choose finite system size of $N=300$ lattice sites and $B_{x}=2.0$ for the topological regime. All the other model parameters are chosen as~$(c_{0},t_{h},u,\Delta)$=(1.0,1.0,0.5,1.0). We consider $\nu=\pi/50$ throughout our manuscript.}
	\label{fig:static_NW}
\end{figure}

To study the transport signature of static MZMs, we begin with topological regime of the static NW Hamiltonian in Eq.~(\ref{eq:Hamiltonian}) hosting one MZM at each of its end. We compute single-terminal differential conductance $\sigma_{\rm{stat}}$ using Eq.~(\ref{eq:main_stat_sigma}) and show with respect 
to the bias voltage $V$~(see Fig.~\ref{fig:static_NW}). We obtain several small peaks outside the bulk gap in addition to a large peak existing at $V=0$, called the ZBP. The latter one exhibits a quantized value of $2e^{2}/h$ as a signature of MZMs, while formers do not show any such universal feature~\cite{KunduPRL2013}. This ZBP emerges due to the resonant Andreev reflection caused by the end localized MZM~\cite{Law_quantization_2009}.


\subsection{Transport signature of single $0$- and/or $\pi$- FMEMs present at one end of the NW}\label{sub:single_NW}
To explore transport signature of FMEMs, we apply step drive protocol as mentioned in Eq.~(\ref{eq:step_drive}) to the Rashba NW setup. The driven system hosts emergent FMEMs~\cite{Mondal2023_NW}. Here, it is worth mentioning the nomenclature for FMEMs: if the driven system hosts total one pair~(one mode at each end) of Majorana modes with quasi-energy $0$~($\Omega/2$), we label them single $0$-~($\pi$)-mode. 

Here, we consider four cases on the basis of number of FMEMs belong to particular quasi-energy, located at one end of the NW and their corresponding bulk gap structure.
Case 1: only one $0$-mode, case 2: only one $\pi$-mode, case 3: one $0$- and one $\pi$-mode with larger bulk gap for $0$-mode as compared to the $\pi$-mode, case 4: same as case 3 with larger bulk gap of $\pi$-mode than the $0$-mode. Note that, Ref.~\cite{Mondal2023_NW} contains rich phase diagram for number of emergent FMEMs in driven Rashba NW setup. Following this, we consider only four representative cases out of many cases to explain the main physical outcome. To compute $\tilde{\sigma}$, we examine Floquet states in frequency space $|u_{\bar{\alpha}}^{(n)}\rangle$. We obtain these states by diagonalizing the extended space Hamiltonian truncated upto 
$n=\pm 10$ photon index~(see~Appendix~\ref{sec:Extended space Hamiltonan} for details). 

\begin{figure*}[]
	\centering
	\subfigure{\includegraphics[width=1.0\textwidth]{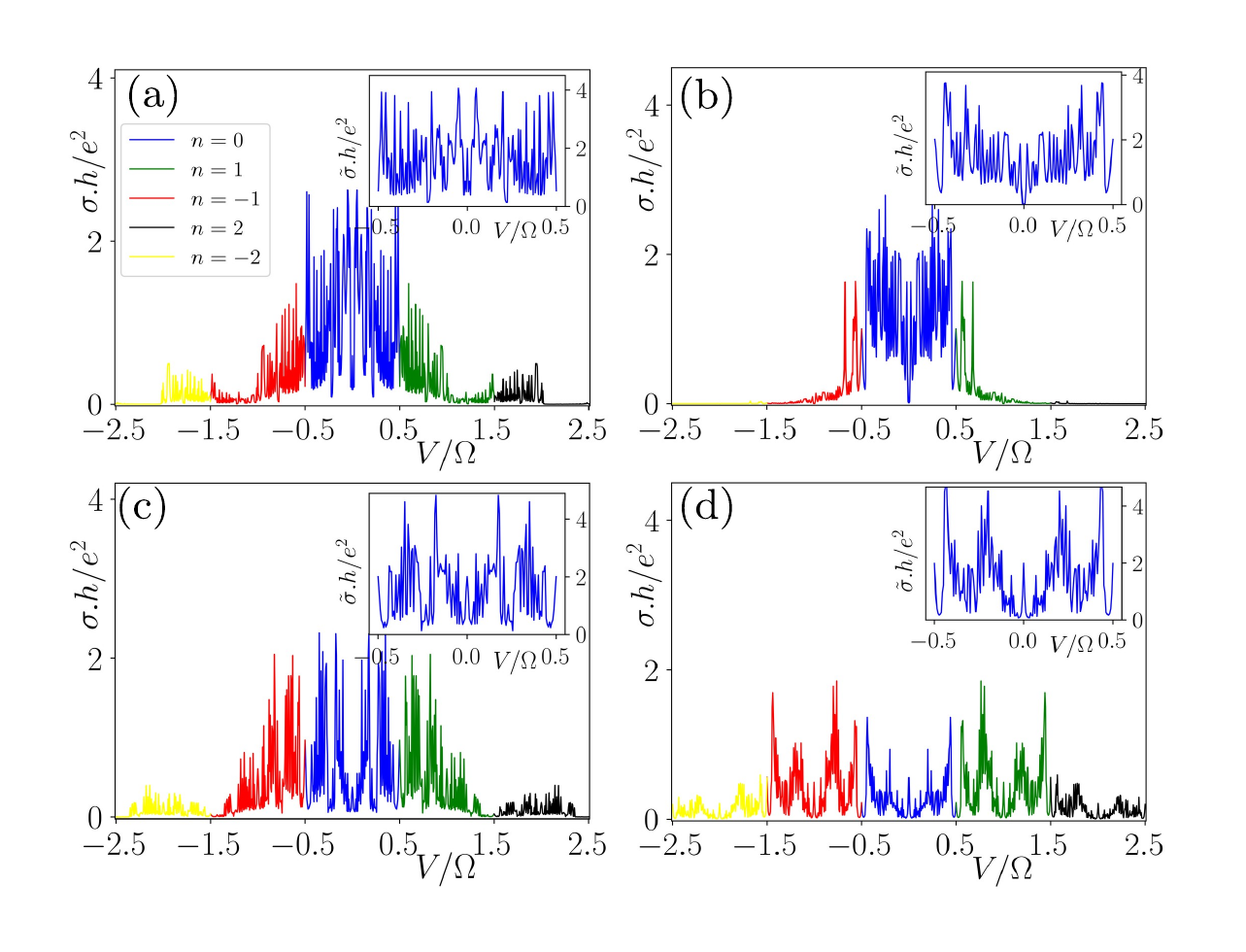}}
	\caption{In case of driven Rashba NW setup, the single-terminal differential conductance $\sigma$ appearing from different photon sectors~(their sum $\tilde{\sigma}$) are depicted in the panels~(insets). Panel (a) [(b)] stands for the same considering single $0$- [$\pi$-] FMEMs present at one end of the NW, representing case 1 [2]. On the other hand, the panels (c) and (d) both correspond to the simultaneous presence of a single $0$-mode and a single $\pi$-mode at one end of the NW, capturing cases 3 and 4, respectively. For all the above cases, no single photon sector contributes alone to the quantized value. However, summed conductance $\tilde{\sigma}$ exhibits a quantize value of $2e^{2}/h$ at $V=0$ and/or $V=\Omega/2$. Here, the model parameters are chosen as (a) $(c_{0},c_{1},\Omega,N)$ = $(0.25,0.25,2.25,110)$, (b) $(c_{0},c_{1},\Omega,N)$ = $(1.0,-1.8,4.5,70)$, (c) $(c_{0},c_{1},\Omega,N)$ = $(-0.45,-0.5,2.24,110)$ and (d) $(c_{0},c_{1},\Omega,N)$ = $(0.8,1.92,1.49,100)$. We choose $B_{x}=1.0$ and all the other model parameters remain same as mentioned in Fig.~\ref{fig:static_NW}.}
	\label{fig:photon_NW}
\end{figure*} 

For all the four cases mentioned above, we compute one-terminal differential conductance $\sigma$ for individual photon sectors, and their sum $\Tilde{\sigma}$ using Eq.~(\ref{eq:main_sum_slpit}). 
We illustrate them in four panels of Fig.~\ref{fig:photon_NW}. In Fig.~\ref{fig:photon_NW}(a) we represent $\sigma$ for case 1 with single 0-mode. It is evident that no photon sector individually exhibits a quantized peak for bias voltage $V+n\Omega$. However, the summed conductance $\tilde{\sigma}$ manifests a quantized peak of \textcolor{red}{$2e^{2}/h$} at $V=0$~[see the inset of Fig.~\ref{fig:photon_NW}(a)]. Then, Fig.~\ref{fig:photon_NW}(b) stands for the transport signature corresponding to case 2 with single $\pi$-mode. Here also $\tilde{\sigma}$ exhibits a quantized peak of $2e^{2}/h$ at $V=\Omega/2$~[see the inset of panel Fig.~\ref{fig:photon_NW}(b)] while contribution originating from the individual photons sectors $\sigma$ exhibits no such quantization. Furthermore, Fig.~\ref{fig:photon_NW}(c) and (d) represent the transport signature for case 3 and case 4, respectively, with one $0$- and one $\pi$-mode. In these above two cases, we also obtain the quantized nature of the  summed conductance $\tilde{\sigma}$ 
at $V=0$ and $V=\Omega/2$. Hence, our results are quite consistent with the well known Floquet sum rule~\cite{KunduPRL2013} even in case of Rashba NW setup. It is important to note that individual photon sector exhibits off-resonant Andreev reflection as well as normal reflection, which gives rise to a non-quantized differential conductance. However, the summed conductance remains quantized, mimicking resonant Andreev reflection for the FMEM. Another noteworthy observation is that the symmetry in the contributions arising from different photon sectors, specifically between the $n^{\rm{th}}$ and $-n^{\rm{th}}$ photon sectors. This symmetry suggests that particle-hole symmetry of the system is preserved within individual photon sectors as well.


\subsection{Stability against disorder }\label{sub:dis}

\begin{figure*}[]
	\centering
	\subfigure{\includegraphics[width=1.0\textwidth]{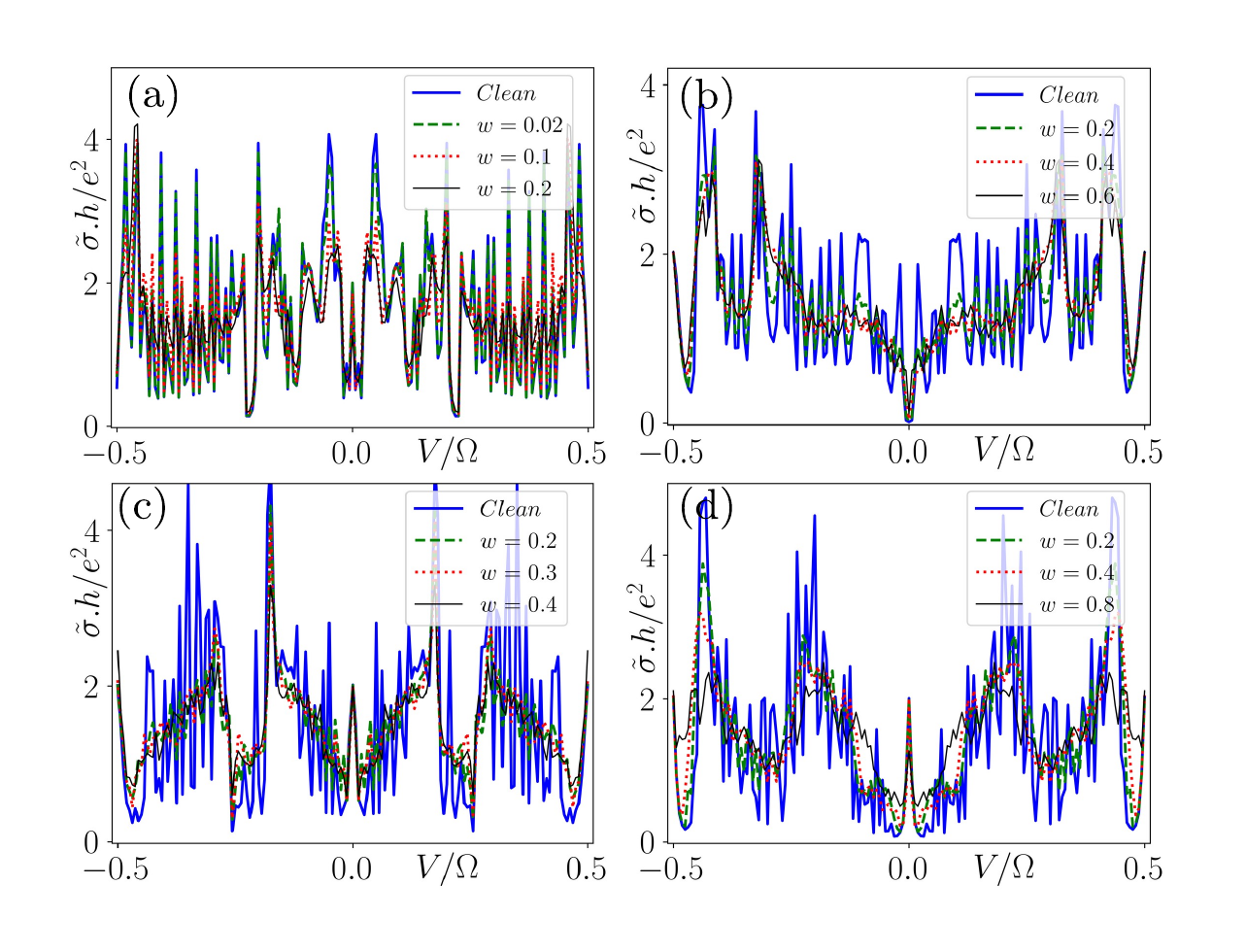}}
	\caption{We depict the disorder averaged one-terminal summed differential conductance $\tilde{\sigma}$ as a function of voltage bias $V$ in the presence of random onsite disorder.  We here consider the same order for panels (a,b,c,d) corresponding to the cases 1, 2, 3, and 4, respectively as presented in Fig.~\ref{fig:photon_NW}. The  value of model parameters remain the same as mentioned in Fig.~\ref{fig:photon_NW}. We consider $100$ disorder configurations for each case.}
	\label{fig:disorder}
\end{figure*}


To investigate the robustness of $\tilde{\sigma}$ for the above mentioned cases, we introduce 
time-independent random onsite chemical potential in the Hamiltonian~[Eq.~(\ref{eq:Hamiltonian})] as
\begin{eqnarray}
	V_{\rm dis}=V(r) \Gamma_{1}\ , \label{eq:dis}
\end{eqnarray}
where $V(r)$ is randomly distributed over the range $\left[-\frac{w}{2},\frac{w}{2}\right]$ with disorder strength given by $w$. Over the full time period of the drive, the lattice Hamiltonian with onsite disorder takes the form as
\begin{eqnarray}
	H_{\rm dis}(t)&=&H_{1}+\sum_{j=1}^{N}\Psi_{j}^{\dagger} \left[ V(r) \Gamma_{1} \right]\Psi_{j}; \hspace*{0.3 cm} t\in \left[0,\frac{T}{4}\right) \nonumber \\ 
	&=& H_{0}+\sum_{j=1}^{N}\Psi_{j}^{\dagger} \left[ V(r) \Gamma_{1} \right]\Psi_{j}; \hspace*{0.3 cm} t\in \left[\frac{T}{4},\frac{3T}{4}\right)\nonumber	\\
	&=& H_{1}+\sum_{j=1}^{N}\Psi_{j}^{\dagger} \left[ V(r) \Gamma_{1} \right]\Psi_{j}; \hspace*{0.3 cm} t\in \left[\frac{3T}{4},T\right] \label{eq:step_dis}.
\end{eqnarray} 
Similar to the clean case as mentioned in the previous subsection, we compute $\tilde{\sigma}$ in the presence of disorder. We consider three disorder strengths for each case and perform an average 
over $100$ disorder configurations. We illustrate our results in Fig.~\ref{fig:disorder}. 

Here, Fig.~\ref{fig:disorder}(a) demonstrates the disorder study for case 1. The quantized value of 
$\tilde{\sigma}$ at $V=0$ is quite stable against weak disorder~($w=0.02$) and starts loosing its quantization in modarate disorder limit~($w=0.1,0.2$). Similarly, Fig.~\ref{fig:disorder}(b) represents the same for case 2. Here also quantization of $\tilde{\sigma}$ at $V=\Omega/2$ remains unaffected 
in weak ~($w=0.2$) and moderate~($w=0.4$) disorder limit, and fails to retain the quantization (\ie peak height decreases) towards strong~($w=0.6$) disorder regime. Note that, weak~(strong) disorder corresponds to its strength being smaller~(larger) than the corresponding bulk gap and moderate disorder refers to the strength equivalent to the corresponding bulk gap. Furthermore, Fig.~\ref{fig:disorder}(c)~(Fig.~\ref{fig:disorder}(d)) describes the case 3~(case 4). In both the instances, weak disorder cannot affect the quantization at $V=0$ and $V=\Omega/2$. However, the scenario becomes different when we start increasing the disorder strength. At first, let us concentrate on Fig.~\ref{fig:disorder}(c). We find that $w=0.3$ destroys the quantization at $V=\Omega/2$ while the peak 
height at $V=0$ is unaffected. One has to increase the disorder strength to $w=0.4$ to diminish its quantization. Therefore, in this case, the ZBP~($V=0$) is more stable against disorder than the
$\pi$ bias~($V=\Omega/2$) peak. Nevertheless, Fig.~\ref{fig:disorder}(d) describes exactly the opposite scenario to the previous case where $\pi$ bias peak is found to be more stable than the 
zero bias one when disorder is present. Nonetheless, for both situations, a more stable peak is associated with a larger bulk gap~(0 or $\pi$) which is expected in the study of disorder.

\subsection{Transport signature for multiple FMEMs}\label{sub:multi}
\begin{figure}[]
	\centering
	\subfigure{\includegraphics[width=0.49\textwidth]{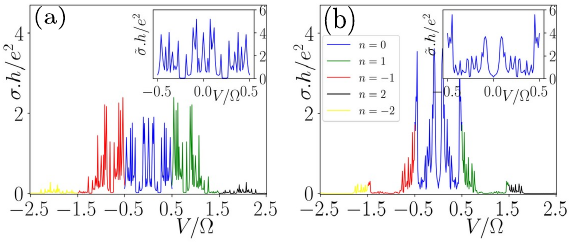}}
	\caption{(a) We depict the one terminal differential conductance $\sigma$ as a function of bias voltage $V$ choosing different photon sectors for double $0$-modes in case of driven Rashba NW. The inset illustrates the quantized peak obtained via summed conductance $\tilde{\sigma}$ for the same setup. (b) Here we repeat the same for $\pi$-modes. Note that, $\tilde{\sigma}$ exhibits a quantized peak value of $4e^{2}/h$ at $0$-($\Omega/2$-) bias voltage for the former~(latter) case. We choose the model parameters as~($c_{0},c_{1},\Omega,N$)=$(0.5,-1.0,2.25,100)$ for panel (a) while $(1.0,0.2,2.4,100)$ for panel (b). We choose $B_{x}=1.0$ and all the other model parameters remain the same as mentioned in Fig.~\ref{fig:static_NW}.}
	\label{fig:multiple}
\end{figure}

After investigating the transport signature of single FMEM corresponding to a particular quasi-energy (0 or $\pi$) and located at one end of the NW, we here discuss the same for multiple FMEMs. 
To be precise, the driven system can host more than one FMEM for a particular quasi-energy at its one end. For completeness, we consider two cases: double $0$- and double $\pi$-modes present at each end of the driven NW. For both the cases, we compute $\sigma$ corresponding to each photon sector and depict them in Fig.~\ref{fig:multiple}(a) and Fig.~\ref{fig:multiple}(b) for $0$- and 
$\pi$-modes, respectively. Here also no individual photon sector gives rise to a quantized peak for $\sigma$. However, their sum exhibits a quantized peak with height $4e^{2}/h$ at $V=0$ for the former case~(double 0-modes) and for the latter~(double $\pi$-modes) at $V=\Omega/2$~(see the insets of Fig.~\ref{fig:multiple}). 


\section{Alternative experimentally feasible model: helical Shiba chain} \label{sec:Shiba}

\begin{figure}[]
	\centering
	\subfigure{\includegraphics[width=0.46\textwidth]{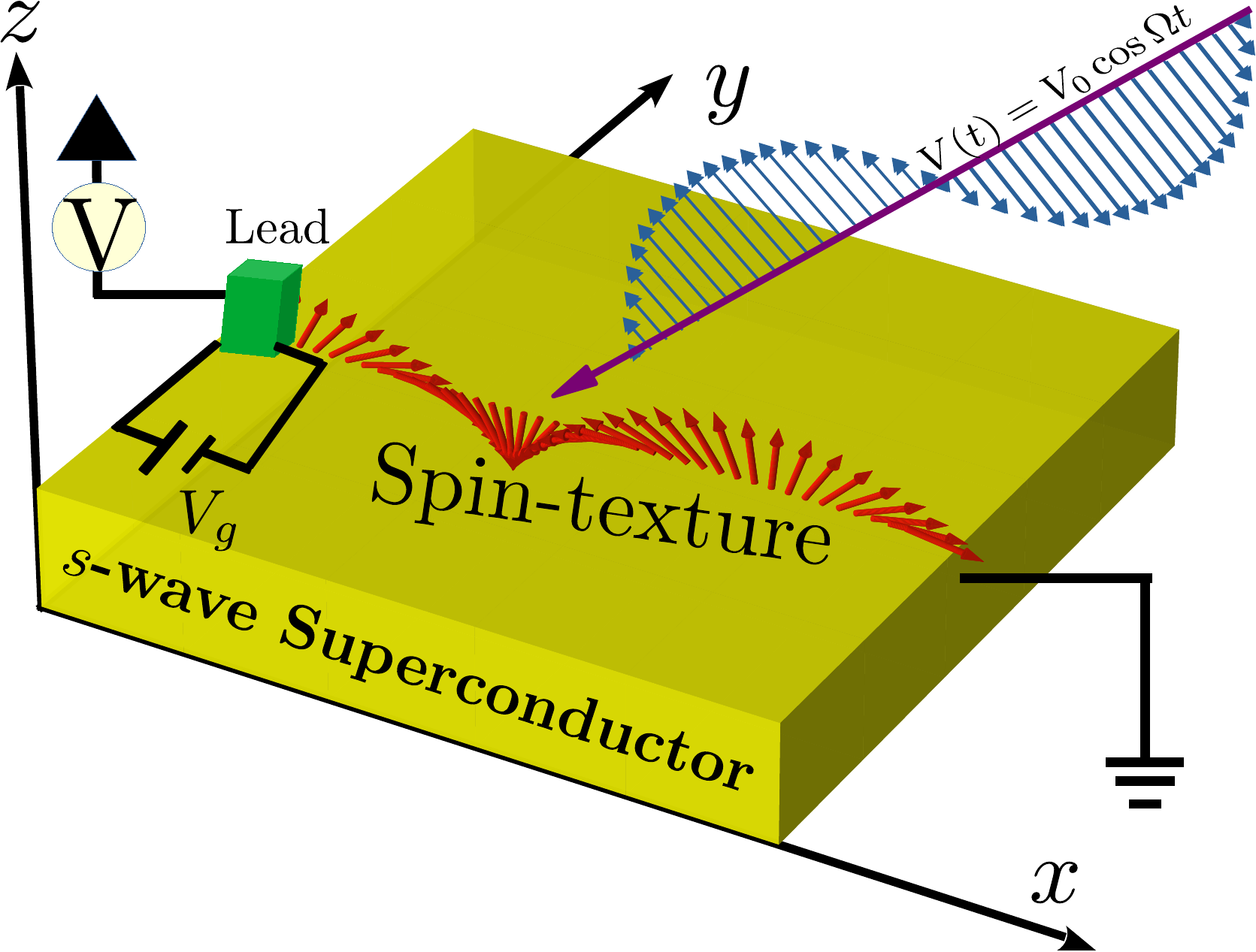}}
	\caption{Schematic diagram of our setup for helical Shiba chain model is depicted. A chain of magnetic adatoms~(red arrows) is placed on the top of a common $s$-wave superconductor~(lime green). The angle between the magnetic moments of two adjacent atoms is such that they form a helical structure~(here out of plane N\'eel type). To change chemical potential a gate voltage $V_{g}$ is applied. A metallic lead~(green) is attached to one end of the chain. A bias voltage $V$ is applied through the lead to study the differential conductance. The superconductor is connected to the ground. Applied periodic drive $V(t)=V_{0} \cos\Omega t$ is also schematically depicted.
	}
	\label{fig:schematic_Shiba}
\end{figure}

After detailed investigations on transport signature of FMEMs in driven Rashba NW model, we concentrate on another realistic model system known as helical Shiba chain which is based on magnetic impurity chain and superconductor heterostructure~\cite{Yazdani2013,Mondal_2023_Shiba}. In this setup, a chain of magnetic impurity atoms~(with their magnetic moments forming a helical structure) is placed on the surface of an $s$-wave superconducting substrate as shown in Fig.~\ref{fig:schematic_Shiba}. The scattering between superconductor electrons and magnetic adatoms forms bound states termed as Shiba states. The hybridyzation among these Shiba states forms Shiba bands. These emerging bands host effective $p$-wave pairing that causes the MZMs to appear at the end of the chain~\cite{Felix_analytics}. Here, the band inversion takes place within the mini gap created between the two lowest Shiba bands leading to a topological phase transition. Note that, the effective $p$-wave gap for this case is long range in nature~\cite{Felix_analytics}. Such long-rangeness of the effective $p$-wave gap in Shiba chain model is distinct over the nearest neighbour effective $p$-wave gap as formed in Rashba NW. 
We begin by considering the following Bogoliubov-de Gennes~(BdG) basis: 
$\Psi_{j}=\left\{c_{j \uparrow},c_{j \downarrow}, c_{j \downarrow}^{\dagger}, -c_{j \uparrow}^{\dagger} \right\}^{\textbf{t}}$; where, $c_{j \uparrow}^{\dagger}$~($c_{j \uparrow}$) and $c_{j \downarrow}^{\dagger}$~($c_{j \downarrow}$) stand for quasi-particle creation (annihilation) operator for the spin-up and spin-down sector at a lattice site-$j$, respectively, and \textbf{t} represents the transpose operation. Then following the above BdG basis, lattice Hamiltonian of a 1D Shiba chain with out-of-plane N\'eel-type spin spiral~\cite{PritamPRB2023} is given by~\cite{Yazdani2013,Mondal_2023_Shiba}
\begin{eqnarray}
		H&=& \sum_{j=1}^{N} \Psi_{j}^\dagger \left[-\mu \Gamma_{1} +B \cos(j \theta) \Gamma_{2} +B \sin(j \theta) \Gamma_{3} + \Delta \Gamma_4 \right] \Psi_{j}\non \\ &&+\sum_{j=1}^{N-1}\Psi_{j,\eta}^\dagger t_{h}  \Gamma_{1} \Psi_{j+1}  +\  {\rm H.c.} 
		\label{eq:Shiba_Hamiltonian} \ ,
\end{eqnarray}
where, $\mu$, $B$, $\theta$, $\Delta$, and $t_{h}$ denote chemical potential, strength of the magnetic impurity, angle between two consecutive classical spins, superconductor gap of the parent 
$s$-wave superconductor and hopping amplitude, respectively. Here, $\Gamma_{1}=\tau_{z} \sigma_{0}$, $\Gamma_{2}=\tau_{0} \sigma_{z}$, $\Gamma_{3}=\tau_{0} \sigma_{x}$, $\Gamma_{4}=\tau_{x} \sigma_{0}$, with the Pauli matrices $\vect{\tau}$ and $\vect{\sigma}$ act on the particle-hole and spin ($\uparrow$, $\downarrow$) sub-spaces, respectively. Note that, the separation between two impurity atoms is large enough to neglect the spin-spin exchange interaction between themselves. The Shiba chain hosts one pair of MZMs at its two ends in the topological superconducting regime which is given by $B_-< \lvert B \rvert < B_+$; with $B_{\pm}=\sqrt{\Delta^{2}+(|\mu|\pm 2 \cos(\theta/2)t_{h})^{2}}$~\cite{Yazdani2013,Mondal_2023_Shiba}. It is interesting to note that $\theta=0$ gives rise to the condition for topological phase transition in case of Rashba NW as mentioned in Sec.~\ref{Sec:II}. Such setup has been experimentally realized in Fe/Co/Mn/Cr adatoms deposited on the top of $s$-wave Nb/Pb/Re superconductor~\cite{Yazdani_science,Yazdani2015,HowonSciAdv2018,Schneider2020,Wiesendanger2021,Beck2021,Wang2021PRL,Schneider2022,Crawford2022}.

Let us consider the following external periodic drive protocol with chemical potential
\begin{eqnarray}
	V(t)= \sum_{j=1}^{N} \Psi_{j}^{\dagger}  \left[ V_{0} \cos (\Omega t) \Gamma_{1} \right] \Psi_{j}  \label{eq:sine_drive}\ ,
\end{eqnarray}
where, $V_{0}$ and $\Omega(= 2\pi/T)$ are the amplitude and frequency of the drive, respectively. The application of such drive gives rise to the topological phase transition within the emergent 
quasi-energy Shiba band hosting FMEMs at the ends of the chain. 
Here, 0 or $\pi$ FMEMs appear even if one starts from the non-topological regime of the system~\cite{Mondal_2023_Shiba}. 

\begin{figure}[]
	\centering
	\subfigure{\includegraphics[width=0.49\textwidth]{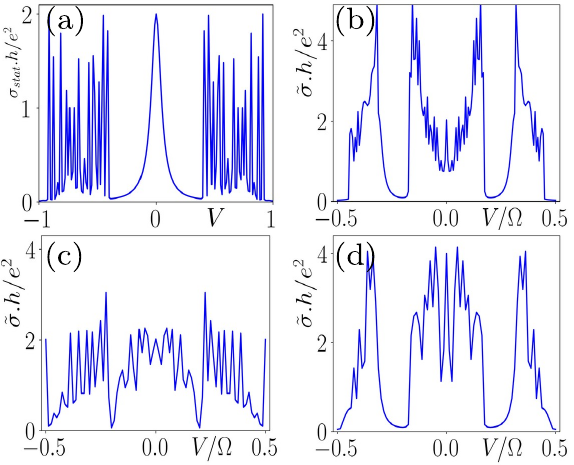}}
	\caption{(a) We depict the one terminal differential conductance $\sigma_{\rm{stat}}$ as a function of bias voltage for the static Shiba chain setup. We obtain a quantized peak with height $2e^{2}/h$ 
at $V=0$. For the driven system we compute conductance $\tilde{\sigma}$~(summed over different photon sectors) considering the cases with single $0$-mode, one $0$- and one $\pi$-modes, and double $0$-modes and illustrate them as a function of bias voltage in panels (b), (c), and (d), respectively. We obtain $2e^{2}/h$ quantization at $V=0$ for panel (b) and at $V=0$ as well as $V=\Omega/2$ for panel (c). In panel (d), ZBP manifests its quantized value of $4 e^{2}/h$ due to double $0$-FMEMs. We choose $B=4$, N=300 lattice sites for panel (a). For panel (b) the model parameters are chosen as 
$(B, N, V_{0}, \Omega) =(5,100,5,7)$ and $(B, N, V_{0}, \Omega) = (2,70,5,6)$ for panel (c). For panel (d), we choose $(B, N, V_{0}, \Omega) =(5,80,8,4.2)$. The rest of the model parameters remain 
the same for all four panels: $(\mu,\Delta,t_{h},\theta)=(4,1,1,2\pi/3)$.}
	\label{fig:Shiba_result}
\end{figure}

First, we compute the static one terminal differential conductance $\sigma_{\rm{stat}}$ in the topological regime of this setup. The ZBP with a quantized value of $2 e^{2}/h$ manifests itself as a transport signature of static Majorana modes as shown in Fig.~\ref{fig:Shiba_result}(a). The other non-quantized peaks~(at non-zero $V$) in $\sigma_{\rm{stat}}$ correspond to the signature of Shiba bands.

For periodically driven Shiba chain, we consider three cases for transport study: single $0$-mode, one $0$- and  one $\pi$-mode, and double $0$-modes that are located at the ends of the chain. We compute Floquet summed conductance $\tilde{\sigma}$ and depict them in Figs.~\ref{fig:Shiba_result}(b), (c), and (d) corresponding to the three cases respectively. We obtain a ZBP with a height of $2e^{2}/h$ as depicted in Fig.~\ref{fig:Shiba_result}(b). The peak height remains the same for $\pi$-modes individually as shown in Fig.~\ref{fig:Shiba_result}(c). On the other hand, Fig.~\ref{fig:Shiba_result}(d) exhibits a ZBP with the quantized value of $4 e^{2}/h$ when two $0$-FMEMs are present. Hence, for all these cases mentioned for the Shiba chain, we obtain similar qualitative results for $\tilde{\sigma}$ as discussed for the case of Rashba NW model.

\section{Transport results employing NEGF technique} \label{sec:NEGF_result}
Having discussed about the transport signature of FMEMs employing an approximate analytical formula~[Eq.~(\ref{eq:main_sum_slpit})], we repeat our findings by applying purely numerical techniques with the help of NEGF. In order to carry that out, we attach another lead to the right end of the system. Then for applied bias $V$, the differential conductance for the driven system is given by~(see Appendix~\ref{sec:app_transport} for details)
\begin{eqnarray}
{\tilde{\sigma}}(V)&=& - 2 \pi e^{2} \!\!\int \!\!d \omega \!\sum_{n}  \mathbf{T}_{LL}^{(n)}\left[f^{\prime L}(\omega)+f^{\prime R}(-\omega)  \right]\non \\
 &+&\!\! 2 \pi e^{2} \!\!\int \!\!d \omega \!\sum_{n} \! \left[\mathbf{T}_{LR}^{(n)} \hspace*{ 1 mm}f^{\prime R}(\omega)\! -\mathbf{T}_{RL}^{(n)} f^{\prime L}(\omega)\!\right]\non\\
 &=& \sum_{n}\sigma(V+n\Omega)\ , \label{eq:sig_NEGF}
\end{eqnarray}
where, $\mathbf{T}_{\lambda \lambda^{\prime}}^{(n)}(\omega)= \operatorname{Tr}\left[\mathbf{G}^{(n)\dagger}(\omega)  \mathbf{V}^{\lambda \mathbf{t}} \mathbf{G}^{(n)}(\omega) \mathbf{V}^{\lambda^{\prime}} \right]$ and $\lambda,\lambda^{\prime} \in L,R$ denote the lead indices. For the static case, the zero frequency limit of the above equation gives rise to the corresponding expression for differential conductance $\sigma_{\rm{stat}}$ as
\begin{eqnarray}
    \sigma_{\rm{stat}}(V)&=& - 2 \pi e^{2} \!\!\int \!\!d \omega~\! \mathbf{T}_{LL}\left[f^{\prime L}(\omega)+f^{\prime R}(-\omega)  \right]\non \\
 &+&\!\! 2 \pi e^{2} \!\!\int \!\!d \omega~\! \! \left[\mathbf{T}_{LR} \hspace*{ 1 mm}f^{\prime R}(\omega)\! -\mathbf{T}_{RL} f^{\prime L}(\omega)\!\right]\ . \non\\
    &&\label{eq:stat_NEGF}
\end{eqnarray}
Here,  $\mathbf{T}_{\lambda \lambda^{\prime}}(\omega)= \operatorname{Tr}\left[\mathbf{G}^{\dagger}(\omega)  \mathbf{V}^{\lambda \mathbf{t}} \mathbf{G}(\omega) \mathbf{V}^{\lambda^{\prime}} \right]$ with $\mathbf{G}(\omega)=\left[ \omega~\mathbf{I} - \mathbf{H} + i \epsilon ~\mathbf{I}\right]^{-1}$ with infinitesimally small positive value of $\epsilon$. 

\begin{figure*}[]
	\centering
	\subfigure{\includegraphics[width=0.98\textwidth]{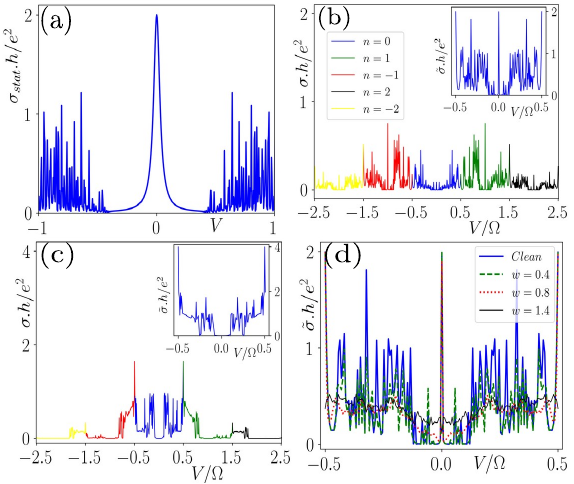}}
	\caption{We depict the two terminal differential conductance as a function of bias voltage $V$ using NEGF technique for the static and driven Rashba NW case. (a) We repeat the inset of Fig.~\ref{fig:static_NW} using Eq.~(\ref{eq:stat_NEGF}) for static case considering $N=120$ lattice sites. We obtain $2 e^{2}/h$ quantization of ZBP for $\sigma_{\rm{stat}}$. Panels (b) and (c) represent the NEGF technique counterpart of Fig.~\ref{fig:photon_NW}(d) and Fig.~\ref{fig:multiple}(b) having one $0$ - one $\pi$-FMEMs and two $\pi$-FMEMs, respectively. We obtain corresponding summed conductance ${\tilde{\sigma}}$ to be quantized at $V=0$ and $V=\Omega/2$ with peak height $2 e^{2}/h$ for single FMEM and $4 e^{2}/h$ for double FMEMs, respectively. (d) We illustrate disorder stability of FMEMs corresponding to panel (b). We observe that for weak disorder limit, quantization of ${\tilde{\sigma}}$ remains unaffected. However, strong disorder destroys the quantization of peak height. We consider all the model parameters 
		to be the same as mentioned before.}
	\label{fig:NW_NEGF}
\end{figure*}

\subsection{Rashba NW}\label{sec:NEGF_NW}
We illustrate our transport results in Fig.~\ref{fig:NW_NEGF} for the Rashba NW using NEGF technique. For static NW, we compute differential conductance $\sigma_{\rm{stat}}$ using Eq.~(\ref{eq:stat_NEGF}) and illustrate as a function of bias voltage $V$ in Fig.~\ref{fig:NW_NEGF}(a). Like previous method, we obtain the ZBP to be quantized with the value $2 e^{2}/h$. 
For the driven Rashba NW case, we compute differential conductance considering individual photon sectors $\sigma$ and their sum ${\tilde{\sigma}}$ using Eq.~(\ref{eq:sig_NEGF}). The corresponding
results are depicted in Fig.~\ref{fig:NW_NEGF}(b) and (c) considering one 0-, one $\pi$-FMEMs and double $\pi$-FMEMs, respectively. Here also akin to the earlier method, no photon sector solely gives rise to the quantized peak for $\sigma$ while their sum ${\tilde{\sigma}}$ exhibits quantization of $2 e^{2}/h$ at $V=0$ and $V=\Omega/2$ for the former case [see the inset of Fig.~\ref{fig:NW_NEGF}(b)] and $4e^{2}/h$ at $V=\pm\Omega/2$ for the latter case as shown in the inset of Fig.~\ref{fig:NW_NEGF}(c). We also present disorder stability of ${\tilde{\sigma}}$ considering one 0-, one $\pi$-FMEMs 
in Fig.~\ref{fig:NW_NEGF}(d). Note that, quantized peak height remains robust against weak disorder strength while looses quantization completely for strong disorder strength. Therefore, our NEGF 
results match well with the results obtained from approximate analytical formula~[Eq.~(\ref{eq:main_sum_slpit})].

\begin{figure}[]
	\centering
	\subfigure{\includegraphics[width=0.5\textwidth]{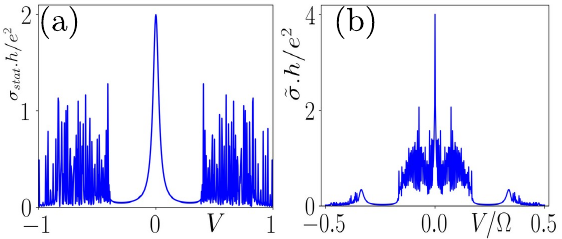}}
	\caption{Here, we depict two terminal differential conductance as a function of bias voltage $V$ using NEGF technique for the static and driven Shiba chain model. In panels (a) and (b), we repeat 
		Fig.~\ref{fig:Shiba_result}(a)~[static case] considering $N=100$ lattice sutes and Fig.~\ref{fig:Shiba_result}(d)~[driven system with double 0-FMEMs] respectively using Eq.~(\ref{eq:stat_NEGF}). 
		We obtain $2 e^{2}/h$~[$4 e^{2}/h$] quantization of ZBP for $\sigma_{\rm{stat}}$~[${\tilde{\sigma}}$] in panel (a)~[(b)]. All the model parameters remain same as mentioned in Fig.~\ref{fig:Shiba_result}. 
	}
	\label{fig:Shiba_NEGF}
\end{figure}
\subsection{Helical Shiba Chain}\label{sec:NEGF_Shiba}
 We depict our NEGF results for the  helical Shiba chain in Fig.~\ref{fig:Shiba_NEGF}. Here, Fig.~\ref{fig:Shiba_NEGF}(a) represents $\sigma_{\rm{stat}}$ for the static Shiba chain while 
 Fig.~\ref{fig:Shiba_NEGF}(b) stands for ${\tilde{\sigma}}$ considering double $0$-FMEMs emerged in driven Shiba chain setup. We obtain ZBP of $\sigma_{\rm{stat}}$~(${\tilde{\sigma}}$) 
 to be quantized with the peak height $2e^{2}/h$~($4e^{2}/h$) for the static MZM~(double $0$- FMEMs). Hence, in case of Shiba chain also we obtain the same results using NEGF method as the earlier case.


\section{Experimental feasibility} \label{sec:experiment}
In this section, we discuss possible experimental connection to realize our numerical findings regarding transport signature of FMEMs. 
\subsection{Rashba NW}
In case of Rashba NW, as far as experimental feasibility is concerned, suitable candidate can be InAs/InSb NW having strong SOC and placed on the top of Nb/Al~($s$-wave superconductor). 
Au can be used  as metallic lead~\cite{das2012zero}. Three step drive protocol [Eq.~(\ref{eq:step_dis})] may be applied via time dependent gate voltage~\cite{Misiorny18,khosravi2009bound,dubois2013minimal,Gabelli13}, by choosing proper superposition of several harmonics. Note that, for $0\leq t <T/4$ and $3T/4\leq t < T$ the system acts as an atomic insulator without any hopping. Hence, in these time domains gate voltage and other control parameters need to be tuned in such a way that the bands remain relatively flat and largely gapped. The band dispersion plays a crucial role only in the middle half of the drive~($T/4\leq t <3T/4$). To obtain possible experimental signature of one $0$- and one $\pi$- FMEMs \eg Fig.~\ref{fig:photon_NW}(d), following parameters may be considered: reported value for Rashba SOC~($2u$) in InSb is around $50~{\rm{\mu eV}}$~\cite{Mourik2012Science}. Our numerical simulation suggests $t_{h}=2u=\Delta=B_{x}=50~{\rm{\mu eV}}$, $c_{1}=0.8t_{h}=40 ~{\rm{\mu eV}}$, $c_{1}=1.92t_{h}= 96~{\rm{\mu eV}}$, and $\Omega=1.49t_{h} \approx 18~{\rm{GHz}}$.

\subsection{Helical Shiba chain}
For Shiba chain model, Nb~(110) can be one good choice with large superconducting gap $\Delta=1.51$ meV~\cite{Schneider2021}. Afterwards, Mn/Cr magnetic adatoms can be fabricated on top of  Nb~(110) substrate by using STM-based single-atom manipulation technique~\cite{HowonSciAdv2018,Schneider2020, Schneider2022,Eigler1990}. This method provides a better tunability for the angles between spin of the impurity atoms~\cite{Yazdani_science}. For this case also, Au can be the suitable choice for metallic lead. External sinusoidal drive can be applied via AC gate voltage. To obtain possible transport signature \eg Fig.~\ref{fig:Shiba_result}(b), our numerical computation suggests the model parameters to have values: $t_{h}=\Delta=1.51~{\rm{meV}}$, $B=2 \Delta=3.02~{\rm {meV}}$, $V_{0}= 5 \Delta=7.55~{\rm{meV}}$, and $\Omega \approx 1.45~{\rm{THz}}$.

\section{Discussions and Outlook}\label{sec:discussion}
It is known in the literature that a single Majorana mode at the vortex core of a $p$-wave superconductor or MZM appeared at the end of the nanowire 
exhibits only resonant Andreev reflection leading to quantized ZBP of $2e^{2} /h$ 
at zero temperature~\cite{Law_quantization_2009, You_JPCM_2015, Tuovinen_NJP_2019,  Prada2020, Gonzalez_2020}. 
Note that, zero bias conductance peak may also arise due to Andreev's bound states (ABSs)~\cite{Prada2020}. However, within the topological regime, enhancement of the strength of the magnetic field can result in splitting of the ABSs, while the MZMs still persist. This subtle issue can 
further be possible to resolve via the shot noice study. The current fluctuation cross correlation called shot noise obeys a universal relationship with applied bias voltage $V$ as ~$-1/V$ for MZMs. Unlike differential conductance, this feature is valid at finite temperature as well~\cite{HaimPRL2015,HaimPRB2015,Mondal_shot2024,Smirnov_PRB_2023}. Although, developing appropriate formalism for shot noise in case of FMEMs is yet to explore. Moreover, Josephson current exhibits $4 \pi$ periodicity with respect to superconducting phase difference between two superconductors for both static MZMs~\cite{Jose_PRL_108,Jose_NJP_2013}, and FMEMs~\cite{Gavensky_PRB_2021,Peng_PRR_2021,KumariPRL}. Therefore, investigation of
shot noise  and Josephson current signature of FMEMs in these realistic models can be interesting future directions as well.

\section{Summary and Conclusions} \label{sec:conclusion}
To summarize, in this article, we explore the transport signature of single and multiple FMEMs in two experimentally feasible setups: first one is Rasha NW and the second one is helical Shiba chain model
based on magnet-superconductor heterostructure. Initially, we begin with a 1D NW, having strong Rashba SOC and Zeeman field oriented along its length, that is placed in close proximity to an $s$-wave superconductor. The system can effectively mimic the 1D $p$-wave Kitaev chain by hosting MZMs at the two ends of the NW. We study the transport signature of these MZMs by computing one-terminal static differential conductance $\sigma_{\rm{stat}}$. We obtain a ZBP with quantized peak height of $2 e^{2}/h$ for $\sigma_{\rm{stat}}$ as the indirect signature of MZMs~\cite{KunduPRL2013}. Thereafter, we adopt a three-step periodic drive protocol in this model to engineer anomalous $\pi$-modes in addition to regular $0$-modes. One can have a control over the number of FMEMs by regulating the frequency and amplitude of the drive~\cite{Mondal2023_NW}. We obtain $ 2e^{2}/h$ quantization in summed conductance for both types of FMEMs in RNW system. 

We also explore the stability of these quantized peaks against static random onsite disorder. We observe that the quantization remains robust in the limit of small disorder strength and the peak height loses its quantization in the limit of moderate or high disorder limit depending on the $0$ or $\pi$ gap. We find that the peaks due to FMEMs associated with larger bulk gaps exhibit better stability against disorder. For multiple FMEMs, we obtain the corresponding quantized value $n_{M}\times 2 e^{2}/h$ with $n_{M}$ being the number of FMEMs that are localized at one end for a large enough system. We repeat our study for helical Shiba chain model~\cite{Yazdani2013,Felix_analytics,Mondal_2023_Shiba} and obtain similar results as compared to the Rashba NW. We further validate our investigation using the microscopic NEGF framework. Here also, we obtain a quantized peak of $n_{M}\times 2e^{2}/h$ for differential conductance corresponding to $n_{M}$ number of FMEMs present at one end of the system. This clearly indicates qualitative match between the analytical and numerical methods as far as quantized conductance is concerned. 
Finally we discuss the experimental feasibility of our findings for both the models. 

Under the application of external periodic drive, the concerned two realistic systems host multiple FMEMs. Braiding among these non-equilibrium Majorana modes can establish themselves as suitable candidates for fault-tolerant topological quantum computations. However, primarily experimental detection of them is necessary before proposing any such kind of application. Transport signature is a very important tool as far as the detection of topological FMEMs is concerned. Hence, our study paves the way towards the detection of multiple FMEMs in realistic model systems. 

\subsection*{Acknowledgments}
We acknowledge Arijit Kundu, Arnob Kumar Ghosh, Amartya Pal, and Kamalesh Bera for stimulating discussions. D.M. thanks Dipak Maity and Sanu Varghese for technical help. D.M. and A.S. also acknowledge SAMKHYA: High-Performance Computing Facility provided by Institute of Physics, Bhubaneswar and the two workstations provided by Institute of Physics, Bhubaneswar from DAE APEX Project, for numerical computations. T.N. acknowledges NFSG from Grant No. BITS Pilani NFSG/HYD/2023/H0911. 

\appendix
\section{Notations}~\label{sec:notations}
Throughout all the appendices, we use the following notations for indices. Greek letters like $\bar{\alpha}$, $\bar{\beta}$ are used to denote the eigenvalue indices and $\alpha$, $\beta$ stand for site indices in the leads. We consider $i$, $j$, $l$ for site indices in the system while $k$, $m$, $n$ indicate frequency indices. Furthermore, $\lambda$, $\lambda^{\prime}$ represent lead indices~(left $L$ 
or right $R$). Superscript index $c$ is used for contact site of the system. The $s$ and $s^{\prime}$ indices represent spin/chiral degrees of freedom. Note that, bold notation in equations corresponds 
to matrix representation and corresponding superscript $T$ stands for transpose operation. Also, repeated indices follow Einstein's summation rule. We consider $\hbar=1$ following the natural unit.
\section{A primer to Floquet theory}~\label{sec:primer_Floquet}
For a periodically driven system~\ie $H(t+T)=H(t)$, Schr{$\ddot{\text{o}}$}dinger's equation can be written as~\cite{Eckardt_2015,Rodriguez-Vega_2018,KunduPRL2013}
\begin{eqnarray}
	i \frac{d}{dt} \ket{\psi(t)} \hspace*{1 mm} = \left(H(t)-i \Sigma \right) \ket{\psi(t)}\ , \label{eq:schrodinger_eqn},
\end{eqnarray}
where $\Sigma$ denotes the environmental degrees of freedom emerging due to the bath attached to the system. This makes the time evolution non-unitary where $T$ is the time period of the drive. Following the analogy of the Bloch theorem, we consider the ansatz as
\begin{eqnarray}
	\ket{\psi_{\bar{\alpha}}(t)} \hspace*{1 mm} = e ^{-i (\epsilon_{\bar{\alpha}} - i \delta_{\bar{\alpha}})t} \ket{u_{\bar{\alpha}}(t)} \label{eq:ansatz}\ ,
\end{eqnarray}
with $\ket{u_{\bar{\alpha}}(t)} \hspace*{1 mm}=\ket{u_{\bar{\alpha}}(t+T)}$. Here, $\ket{u_{\bar{\alpha}}(t)}$ are called Floquet states and $\epsilon_{\bar{\alpha}}$ are called quasi-energies with width $\delta_{\bar{\alpha}}$. Inserting the following ansatz in Eq.~(\ref{eq:schrodinger_eqn}) we have
\begin{eqnarray}
	\left(H(t)-i \Sigma - i \frac{d}{dt} \right) \ket{u_{\bar{\alpha}}(t)} \hspace*{1 mm}= (\epsilon_{\bar{\alpha}} - i \delta_{\bar{\alpha}}) \ket{u_{\bar{\alpha}}(t)} \label{eq:F_S_eqn}.
\end{eqnarray}
The above equation is called Floquet-Schr{$\ddot{\text{o}}$}dinger's equation. Time evolution operator is given by 
\begin{eqnarray}
	U(t,t^{\prime})=\text{TO}\left[ \hspace*{1mm}  e^{-i \int_{t^{\prime}}^{t}H(s) ds} \right] \label{eq:evolution_definition}.
\end{eqnarray}
Note that, for a periodically driven system, time evolution operator is also periodic \ie $U(t+T,T)=U(t,0)$.
Then, $U(t+T,t)\ket{\psi(t)} \hspace*{1 mm}= \ket{\psi(t+T)}$. Using Eq.~(\ref{eq:ansatz}) and periodicity of $|u_{\bar{\alpha}}(t)\rangle$, we have
\begin{eqnarray}
	 U(t+T,t)  \ket{u_{\bar{\alpha}}(t)}  &&= e ^{-i (\epsilon_{\bar{\alpha}} - i \delta_{\bar{\alpha}})T} \ket{u_{\bar{\alpha}}(t)} \label{eq:Flo_evo}.
\end{eqnarray}
For a non-interacting system, if $\Lambda_{\bar{\alpha}}$ are the energy eigenvalues of $U(T,0)$, then the quasi-energies are given by $\epsilon_{\bar{\alpha}}=\frac{i}{T} \ln(\Lambda_{\bar{\alpha}})$. 
Note that, in Eq.~(\ref{eq:Flo_evo}) if we substitute $\epsilon_{\bar{\alpha}} \rightarrow \epsilon_{\bar{\alpha}} + n \Omega$ with $n \in \mathbb{Z}$, and $\Omega=2 \pi/T$, then we obtain new 
$\ket{u_{\bar{\alpha}}(t)} \hspace*{ 1mm} \rightarrow \hspace*{1mm} e^{i n \Omega t} \ket{u_{\bar{\alpha}}(t)}$ for the same set of $\{\ket{\psi_{\bar{\alpha}}(t)}\}$. Thus quasi-energies for Floquet states are not unique and are connected via external periodic drive by emission and absorption of virtual photons. Due to this reason we call these frequency indices as photon sectors as long as Floquet theory is concerned. Although quasi-energies are unique within 1\textsuperscript{st} Floquet zone or $0$\textsuperscript{th} photon sector: $-\Omega/2 \leq \epsilon_{\bar{\alpha}} \leq \Omega/2$.
Since the Floquet states are time periodic, we can write them in terms of Fourier modes as
\begin{eqnarray}
	&\ket{u_{\bar{\alpha}}(t)}& \hspace*{ 1 mm} = \sum_{n} e^{-i n \Omega t} \ket{u_{\bar{\alpha}}^{(n)}}, \nonumber\\
	 \text{with} &\ket{u_{\bar{\alpha}}^{(n)}}& = \int_{0}^{T}\frac{dt}{T} e^{i n \Omega t} \ket{u_{\bar{\alpha}}(t)}. \label{eq:basis_transform}
\end{eqnarray}
One can perform the same for time periodic total Hamiltonian, $H_{\rm tot}=H(t)-i\Sigma$. Then in frequency space, Floquet Schr{$\ddot{\text{o}}$}dinger equation appears as
\begin{equation}
	H_{\rm tot}^{(k-n)}|u_{\bar{\alpha}}^{(n)}\rangle = (\epsilon_{\bar{\alpha}} -i \delta_{\bar{\alpha}}+k \Omega  ) |u_{\bar{\alpha}}^{(k)}\rangle\ , \label{Eq:extendedspace}
\end{equation}
with $H_{\rm tot}^{(k-n)}= \int_{0}^{T}\frac{dt}{T} e^{i (k-n)\Omega t} H_{\rm tot}(t) $. Here, the time evolution is non-unitary due to the presence of $\Sigma$. Hence, we have to deal with bi-orthogonalization of non-Hermitian system. The left eigenvectors $\ket{u_{\bar{\alpha}}^{+}(t)}$ is given by
\begin{eqnarray}
	\left(H(t)+i \Sigma - i \frac{d}{dt} \right) \ket{u_{\bar{\alpha}}^{+}(t)} \hspace*{1 mm}= (\epsilon_{\bar{\alpha}} + i \delta_{\bar{\alpha}}) \ket{u_{\bar{\alpha}}^{+}(t)} \label{eq:F_S_eqn_conj}.
\end{eqnarray}
The eigenstates of $U(T,0)$ yields the bi-orthogonalization and completeness relations as
\begin{eqnarray}
	\frac{1}{T} \int_{0}^{T} dt \langle u_{\bar{\alpha}}^{+}(t)|u_{\bar{\beta}}(t) \rangle &=& \delta_{\bar{\alpha} \bar{\beta}}\ ,\nonumber \\
	 \text{and}~~\sum_{\bar{\alpha}} |u_{\bar{\alpha}}(t) \rangle \langle u_{\bar{\alpha}}^{+}(t) | &=&\mathbf{I} \label{eq:comp_t}\ .
\end{eqnarray}
In frequency space these two relations have the form as
\begin{eqnarray}
	\sum_{k} \langle u_{\bar{\alpha}}^{+(k)} |u_{\bar{\beta}} ^{(k)} \rangle &=&\delta_{\bar{\alpha} \bar{\beta}}\ , \nonumber\\
	\text{and}~~\sum_{\bar{\alpha},k} |u_{\bar{\alpha}}^{(k)} \rangle \langle u_{\bar{\alpha}}^{+(k)}|&=&\mathbf{I} \label{eq:comp_fr} \ .
\end{eqnarray}
In this scenario, the time evolution operator is given by
\begin{eqnarray}
	U(t,t^{\prime})=\sum_{\bar{\alpha}} e^{-i(\epsilon_{\bar{\alpha}} -i \delta_{\bar{\alpha}})(t-t^{\prime})}|u_{\bar{\alpha}}(t)\rangle \langle u_{\bar{\alpha}}^{+}(t^{\prime})| \label{eq:U_exp}\ .
\end{eqnarray}
Note that, $U(t,t^{\prime})$ is the two point correlation function in the time domain \ie retarded Green's function $G(t,t^{\prime})$ in the context of dynamics of a system. We use this in the next section to calculate the conductance for a driven system.
\section{Transport theory for Floquet Majorana modes}~\label{sec:app_transport}
In this Appendix, we elaborate on the discussion presented in Sec.~\ref{sec:tr_fl_th} of the main text. The detailed derivations of summed differential conductance for FMEMs are 
described here. Let us consider a periodically driven system attached to the leads as mentioned before. The Hamiltonian corresponding to the system, leads and coupling between them can be written as~\cite{KunduPRL2013,Pereg_PRL_2015,Pereg_PRB_2016,MitraPRB2019,KOHLER2005379}
\begin{eqnarray}
	&H_{s}(t)&=\sum_{i,j} H_{ij}(t) c_{i}^{\dagger} c_{j} \equiv \mathbf{c}^{\dagger} \mathbf{H}(t) \mathbf{c} \label{sys_ham}\ ,\label{eq:sys_ham} \\
	  &H_{L/R}&=\sum_{\lambda} H_{L/R}^{\lambda}=\sum_{\lambda, \alpha, \beta}  F_{\alpha \beta}^{\lambda} a_{\alpha}^{\lambda \dagger} a_{\beta}^{\lambda} \non\\
	  &&\equiv \mathbf{a^{\dagger}}^{\lambda}\mathbf{F}^{\lambda} \mathbf{a}^{\lambda}\ , \label{eq:res_ham}\\
	&H_{c}&=\sum_{\lambda} H_{c}^{\lambda} =\sum_{\lambda, \alpha,i} K_{\alpha, i}^{\lambda} a_{\alpha}^{\lambda \dagger} c_{i} + h.c. \non\\
	&& \equiv \sum_{\lambda} \mathbf{a^{\lambda \dagger}} \mathbf{K^{\lambda}} \mathbf{c} + h.c.\ .\label{eq:coup_ham}
\end{eqnarray}
Here, $a/ c$~($a^{\dagger} /c^{\dagger}$) denote the electronic annihilation~(creation) operators for the lead/system, respectively. Also, $\lambda$ stands for the $\lambda$\textsuperscript{th} lead at left/right sides of the system. The sum of the above three Hamiltonians can be written as $H_{\rm{Total}}(t)=H_{s}(t)+H_{c}+H_{L/R}$.

Heisenberg equation of motion~(EOM) for lead operator $a_{\alpha}^{\lambda}$ is given by
\begin{eqnarray}
	\dot{a}_{\alpha}^{\lambda}&=&-i \left[a_{\alpha}^{\lambda}, H_{\text{Total}}(t) \right] \nonumber\\
	&=& -i \left[a_{\alpha}^{\lambda}, H_{s}(t)+H_{c} + H_{L/R} \right] \nonumber\\
	&=&-i F_{\alpha \beta}^{\lambda} a_{\beta}^{\lambda} - i K_{\alpha l}^{\lambda} c_{l} \label{eq:eom_lead} \ . 
\end{eqnarray}
Similarly, EOM for the system electronic operator $c_{j}$ is given by
\begin{equation}
	\dot{c}_{j}= -i H_{jl}(t) c_{l} -i K_{\alpha j}^{\lambda *} a_{\alpha}^{\lambda} \label{eq:eom_system}\ .
\end{equation}
Rewriting Eq.~(\ref{eq:eom_lead}), we obtain
\begin{equation}
	\left(	i \delta_{\alpha \beta} \frac{d}{dt} - F_{\alpha \beta}^{\lambda} \right) a_{\beta}^{\lambda} = K_{\alpha l}^{\lambda} c_{l}\ . \label{eq:rewrite_eom_lead}
\end{equation}
The Green's function $g_{\beta \nu}^{\lambda}(t,t^{\prime}) =\langle \left[ a_{\beta}^{\lambda}(t)a_{\nu}^{\lambda \dagger}(t^{\prime}) \right] \rangle$ for the leads can be derived from the above equation and is given by
\begin{eqnarray}
	g_{\beta \nu}^{\lambda}(t,t^{\prime}) 
	&&=-i\left[  e^{-i F^{\lambda}(t-t^{\prime})} \theta(t-t^{\prime}) \right]_{\beta \nu} \label{eq:lead_gf} \\
	&& \equiv \left[g^{\lambda}(t-t^{\prime}) \right]_{\beta \nu} \nonumber.
\end{eqnarray}
The solution of Eq.~(\ref{eq:rewrite_eom_lead}) is given by $a_{\alpha}^{\lambda}(t)=\text{complementary function~(CF)} + \text{particular integral~(PI)}$ \ie
\begin{eqnarray}
	a_{\alpha}^{\lambda}(t) &=& i \left[g^{\lambda}(t-t_{0}) \right]_{\alpha \beta } a_{\beta}^{\lambda}(t_{0}) \non\\
	&&+ \int_{t_{0}}^{t} dt^{\prime} \left[g^{\lambda}(t-t^{\prime}) \right]_{\alpha \beta } K_{\beta l}^{\lambda} c_{l}(t^{\prime})\ , 
	 \label{eq:sol_a}
\end{eqnarray}
where $t_{0}$ is the initial time which can be considered as distant past \ie $-\infty$.

With the help of  Eq.~(\ref{eq:sol_a}), we can rewrite Eq.~(\ref{eq:eom_system}) as
\begin{eqnarray}
	&&\left[i \delta_{jl} \frac{d}{dt} - H_{jl}(t) \right] c_{l}(t)\non\\
	&&- i \int_{-\infty}^{t} dt^{\prime} \Gamma_{jl}(t-t^{\prime}) c_{l}(t^{\prime}) = h_{j}(t)
	 \label{eq:with_all_definition}\ ,
\end{eqnarray}
where, the coupling $\Gamma_{jl}$, and the noise $h_{j}$ in the lead can be defined as
\begin{eqnarray}
	\Gamma_{jl}(t-t^{\prime}) &&= -i K_{\alpha j}^{\lambda *} \left[g(t-t^{\prime}) \right]_{\alpha \beta} K_{\beta l}^{\lambda} \label{eq:Gamma_definition}\ , \\
	h_{j}(t) &&= \sum_{\lambda} h_{j}^{\lambda}(t)\nonumber\\
	&&=i K_{\alpha j}^{\lambda *} \left[g(t-t_{0}) \right]_{\alpha \beta} a_{\beta}^{\lambda}(t_{0}) 
 \label{eq:noise_definition}\ .
\end{eqnarray}
The thermal correlation  is defined as 
\begin{eqnarray}
	\braket{h_{j}^{\lambda \dagger}(t) h_{l}^{\lambda^{ \prime}}(t)}= K_{\alpha j}^{\lambda} g_{\alpha \beta}^{\lambda *}(t) K_{\gamma l}^{\lambda^{\prime}*} g_{\gamma \delta}^{\lambda^{\prime}}(t^{\prime}) \braket{a_{\beta}^{\lambda \dagger}(t) a_{\delta}^{\lambda^{\prime}}(t^{\prime})} \non \\ \label{eq:thermal_corr_def}\ ,
\end{eqnarray}
We can write lead Green's function in terms of lead energy eigenbasis as
\begin{eqnarray}
	g_{\mu \nu}^{\lambda}(t)= -i \theta(t) \sum_{\epsilon^{\lambda}} \langle \mu|\epsilon^{\lambda} \rangle \langle \epsilon^{\lambda} | \nu \rangle e^{-i \epsilon^{\lambda}t} \label{eq:lead_eigen}\ .
\end{eqnarray}
Let us define density of states matrix in frequency space as
\begin{eqnarray}
	\rho_{\gamma \alpha}^{\lambda} (\omega) = \sum_{\epsilon^{\lambda}} \langle \gamma|\epsilon^{\lambda} \rangle \langle\epsilon^{\lambda} | \alpha \rangle \delta(\omega-\epsilon^{\lambda}) \label{eq:dos_matrix}\ .
\end{eqnarray} 
Then thermal correlation in frequency space is given by
\begin{eqnarray}
	\braket{h_{j}^{\lambda \dagger}(\omega) h_{l}^{\lambda^{ \prime}}(\omega^{\prime})}\!\!\!&=&\!\!\!(2\pi)^2 \! \rho_{\gamma \alpha}^{\lambda}(\omega) K_{\alpha j}^{\lambda} K_{\gamma l}^{\lambda^{\prime}*} \delta_{\lambda \lambda^{\prime}} \delta(\omega - \omega^{\prime})  \nonumber \\
    && \times f(\omega, \mu_{\lambda}, \!T_{\lambda}) \nonumber \\
	&\equiv& \!\!\!(2\pi)^2 \mathbf{V}_{jl}^{T} \delta_{\lambda \lambda^{\prime}} \delta(\omega - \omega^{\prime}) f(\omega, \mu_{\lambda}, T_{\lambda}) \ ,\nonumber\\
 &&\label{eq:ther_cor_freq_spaace}
\end{eqnarray}
where, $\mathbf{V}^{\lambda}=\mathbf{K}^{\lambda \dagger} \rho^{\lambda} \mathbf{K}^{\lambda}$ and $f(\omega, \mu_{\lambda}, T_{\lambda}) $ is the Fermi distribution function of the reservoir.
Here, $\mu_{\lambda}$, $T_{\lambda}$ denote the chemical potential and temperature of the reservoirs, respectively.

Afterwards, changing the variable $\tau=t-t^{\prime}$ in Eq.~(\ref{eq:with_all_definition}), one can obtain
\begin{eqnarray}
	\left[i \delta_{jl} \frac{d}{dt} - H_{jl}(t) \right] c_{l}(t) - i \int_{0}^{\infty} d\tau \Gamma_{jl}(\tau) c_{l}(t-\tau) = h_{j}(t)\ . \non\\
	 \label{eq:variable_changed}
\end{eqnarray}
Introducing the Green's function $G_{ij}(t,t^{\prime})=\langle c_{i}(t)c_{j}^{ \dagger}(t^{\prime})\rangle$, in matrix notation we have
\begin{eqnarray}
	\left[i \mathbf{I} \frac{d}{dt} - \mathbf{H}(t) \right] \mathbf{G}(t,t^{\prime}) - &&i \int_{0}^{\infty} d\tau \mathbf{\Gamma}(\tau) \mathbf{G}(t-\tau,t^{\prime})\non\\ 
	&& = \delta(t-t^{\prime}) \mathbf{I}\ . 
	 \label{eq:with_GF}
\end{eqnarray}
For any arbitrary $t$, and $t^{\prime}$ the above equation satisfies $\mathbf{G}(t,t^{\prime})=\mathbf{G}(t+T,t^{\prime}+T)$. Hence, we can perform Fourier transform~(FT) for both the $t$, 
and $t^{\prime}$. At first, let's perform the FT with respect to $t^{\prime}$ and we have
\begin{eqnarray}
	\mathbf{G}(t,\omega)&=&\int_{-\infty}^{t} dt^{\prime} \mathbf{G}(t,t^{\prime}) e^{i \omega (t-t^{\prime})} \label{eq:G(t,w)1}\\
	&=& \int_{0}^{\infty} d\tau \mathbf{G}(t,t-\tau) e^{i \omega \tau}\label{eq:G(t,w)2}\ .
\end{eqnarray}
Then, performing FT with respect to $t$, we obtain
\begin{eqnarray}
	\mathbf{G}^{(k)}(\omega) =\frac{1}{T} \int_{0}^{T}dt e^{ik\Omega t} \mathbf{G}(t,\omega)\ ,
\end{eqnarray}
which yields inverse FT as
\begin{eqnarray}
	\mathbf{G}(t,\omega)=\sum_{k} e^{-ik\Omega t}G^{(k)}(\omega)\ .
\end{eqnarray}
Inverting Eqs.~(\ref{eq:G(t,w)1}) and (\ref{eq:G(t,w)2}) we have
\begin{eqnarray}
	\mathbf{G}(t,t-\tau)&=&\frac{1}{2 \pi} \int_{-\infty}^{\infty} d \omega \mathbf{G}(t,\omega) e^{-i \omega \tau} \label{eq:invert1}\ ,\\
	\mathbf{G}(t,t^{\prime})&=&\frac{1}{2 \pi} \int_{-\infty}^{\infty} d \omega \mathbf{G}(t,\omega) e^{-i \omega (t-t^{\prime})} \label{eq:invert2}.
\end{eqnarray}
\begin{widetext}
Using these two equations, Eq.~(\ref{eq:with_GF}) becomes
\begin{eqnarray}
	\left[i \mathbf{I} \frac{d}{dt} - \mathbf{H}(t) \right] \int_{-\infty}^{\infty} \frac{d \omega}{2 \pi} \mathbf{G}(t,\omega) e^{-i \omega (t-t^{\prime})} - i \int_{0}^{\infty} d\tau \mathbf{\Gamma}(\tau) \int_{-\infty}^{\infty} \frac{d \omega}{2 \pi} \mathbf{G}(t-\tau,\omega) e^{-i \omega (t-\tau-t^{\prime})}  &=& \delta(t-t^{\prime}) \mathbf{I} \nonumber\\
	\implies \int_{-\infty}^{\infty} \frac{d \omega}{2 \pi}e^{-i \omega (t-t^{\prime})}\left[i \mathbf{I} \frac{d}{dt}+\omega - \mathbf{H}(t) \right] \mathbf{G}(t,\omega)  - i \int_{0}^{\infty} d\tau \mathbf{\Gamma}(\tau) \int_{-\infty}^{\infty} \frac{d \omega}{2 \pi} \mathbf{G}(t-\tau,\omega) e^{-i \omega (t-\tau-t^{\prime})}  &=& \int_{0}^{\infty} \frac{d \omega}{2 \pi} e^{-i \omega (t-t^{\prime})} \mathbf{I} \nonumber\ .
\end{eqnarray}
Comparing both sides, we obtain
\begin{eqnarray}
	\left[i \mathbf{I} \frac{d}{dt}+\omega - \mathbf{H}(t) \right] \mathbf{G}(t,\omega)- i \int_{0}^{\infty} d\tau \mathbf{\Gamma}(\tau)  \mathbf{G}(t-\tau,\omega) e^{i \omega \tau}=\mathbf{I}\ . \label{eq:comp}
\end{eqnarray}
Note that, here we assume $t>0$, and ${\rm{lim}}_{t^{\prime}\rightarrow t_{-}} G(t,t^{\prime})=0$.
\end{widetext}

Therefore, the solution for electronic operators of the system is given by
\begin{eqnarray}
	c_{l}(t)&=&\int_{-\infty}^{t}dt^{\prime} \left[G(t,t^{\prime})\right]_{lj} h_{j}(t^{\prime}) \nonumber \\
	&=&\int  \frac{d \omega}{2 \pi}  G_{lj}(t,\omega) h_{j}(\omega) e^{-i\omega t} \label{eq:sol for c} \\
	&=&\int  \frac{d \omega}{2 \pi}  e^{-i\omega t} e^{-i k \Omega t} G_{lj}^{(k)} (\omega) h_{j}(\omega)\ , \label{eq:sol for c_2}
\end{eqnarray}
where, $\mathbf{G}^{(k)}(\omega)$ is called the Nambu-Gorkov Green’s function and is given by~\cite{MitraPRB2019}
\begin{eqnarray}
	\mathbf{G}^{(k)}(\omega)&=& \sum_{\bar{\alpha} ,n} \frac{|u_{\bar{\alpha}}^{(k+n)}\rangle \langle u_{\bar{\alpha}}^{+(n)}|}{\omega-\epsilon_{\bar{\alpha}}-n\Omega +i\delta_{\bar{\alpha}}}\ . \label{eq:NEGF}
\end{eqnarray}
Number operator for electron in $\lambda$\textsuperscript{th} lead is, $N^{\lambda}(t)= a_{\alpha}^{\lambda \dagger}(t) a_{\alpha}^{\lambda}(t)$. Hence, Heisenberg EOM yields $\dot{N}^{\lambda}(t)=-i [N^{\lambda}(t),H_{{\rm{Total}}}] $. The net current flowing accross the $\lambda$\textsuperscript{th} contact into the system can be given by
\begin{eqnarray}
	J^{\lambda}(t)&=&(-e) \times(-\dot{N}^{\lambda}(t)) \nonumber \\
	&=& -ie \left[N^{\lambda}(t),H_{\text {Total}} \right] \nonumber \\
	&=& -ie \left( K_{\alpha l}^{\lambda} a_{\alpha}^{\lambda \dagger}(t)c_{l}(t)- K_{\alpha l}^{\lambda *} c_{l}^{\dagger}(t) a_{\alpha}^{\lambda}(t)\right). \label{eq:raw_current}
\end{eqnarray}
Taking average over the lead states, one can obtain
\begin{eqnarray}
	\langle J^{\lambda}(t)\rangle &=& 2 e \operatorname{Im} \left[K_{\alpha l}^{\lambda} \langle a_{\alpha}^{\lambda \dagger}(t) c_{l}(t) \rangle \right]. \nonumber 
\end{eqnarray}
Using the solution of $a^{\lambda}_{\alpha}(t)$ from Eq.~(\ref{eq:sol_a}) and with the help of Eq.~(\ref{eq:noise_definition}) we can write
\begin{eqnarray}
	\langle J^{\lambda}(t)\rangle &=& 2 e \operatorname{Im}\left[ \langle h_{l}^{\lambda \dagger}(t) c_{l}(t) \rangle\right]  \non\\
	&&+ 2 e \operatorname{Im} \left[\int_{- \infty}^{t} dt^{\prime} K_{\alpha l}^{\lambda} g_{\alpha \beta}^{\lambda*}(t-t^{\prime})K_{\beta j}^{\lambda*} \langle c_{j}^{\dagger}(t^{\prime})c_{l}(t) \rangle \right] \non\\
	&=&\langle J_{1}^{\lambda}(t)\rangle + \langle J_{2}^{\lambda}(t)\rangle\ . 
	\label{eq:raw_current_2}
\end{eqnarray}
Analysis of the first term~$\la J_{1}^{\lambda}(t)\ra$: Using the solution of $c_{l}(t)$ from Eq.~(\ref{eq:sol for c}), and transforming to the frequency space with the help of  Eq.~(\ref{eq:ther_cor_freq_spaace}) we obtain 
\begin{eqnarray}
	\langle J_{1}^{\lambda}(t) \rangle&=& 2 e \int d \omega \hspace*{1 mm} f^{\lambda}(\omega) \operatorname{Tr}\left[\left(\frac{\mathbf{G}(t,\omega)-\mathbf{G}^{\dagger}(t,\omega)}{2i} \right) \mathbf{V}^{\lambda}(\omega)\right]. \non
\end{eqnarray}
\begin{widetext}
Then taking Hermitian conjugate of Eq.~(\ref{eq:comp}), and changing the variable $\omega$ to $\omega^{\prime}$ one can obtain,
\begin{eqnarray}
	\mathbf{G}^{\dagger}(t,\omega^{\prime})\left[-i \mathbf{I} \frac{d}{dt} +\omega^{\prime} -\mathbf{H}(t) \right] + i \int_{0}^{\infty} d \tau e^{-i \omega^{\prime}\tau}\mathbf{G}^{\dagger}(t-\tau,\omega^{\prime}) \mathbf{\Gamma}^{\dagger}(\tau) =\mathbf{I} \label{eq:com_dag}\ .
\end{eqnarray}
Operating $-\mathbf{G}^{\dagger}(t,\omega^{\prime})\times$ Eq.~(\ref{eq:comp}) + Eq.~(\ref{eq:com_dag})$\times \mathbf{G}(t,\omega)$ yields the following equation mentioned below
\begin{equation}
	\frac{\mathbf{G}(t,\omega)-\mathbf{G}^{\dagger}(t,\omega)}{2i}=- \frac{1}{2} \mathbf{I} \frac{d}{dt}\left[ \mathbf{G}^{\dagger}(t,\omega) \mathbf{G}(t,\omega)\right] + \int_{0}^{\infty} d\tau \operatorname{Re}\left[\mathbf{G}^{\dagger}(t,\omega) e^{i  \omega \tau} \mathbf{\Gamma}(\tau) \mathbf{G}(t-\tau,\omega)\right]~\label{eq:com_dag_new}.
\end{equation}
Let us decompose $\la J_{1}^{\lambda}(t)\ra$ into two parts:
\begin{eqnarray}
    \la J_{1}^{\lambda}(t)\ra&=&\la J_{11}^{\lambda}(t)\ra+\la J_{12}^{\lambda}(t)\ra ~\label{eq:decom_J1}\\
    &=&{\text{I}} +{\text{II}} \non\ ,
\end{eqnarray}
Using Eq.~(\ref{eq:com_dag_new}), we can write the first part~($\text{I}$) as
\begin{eqnarray}
	\langle J_{11}^{\lambda}(t) \rangle &=& - e \frac{d}{dt}\int d\omega \hspace*{2 mm}f^{\lambda}(\omega) \operatorname{Tr}\left[\mathbf{G}^{\dagger}(t,\omega)\mathbf{G}(t,\omega) \mathbf{V}^{\lambda}(\omega) \right] \nonumber \\
	\implies \langle \langle J_{11}^{\lambda}\rangle \rangle &=& \frac{1}{T} \int_{0}^{T} dt \hspace*{2 mm} \langle J_{11}^{\lambda}(t) \rangle \nonumber \\
	\implies\langle \langle J_{11}^{\lambda}\rangle \rangle &=& -e \int d\omega \hspace*{2 mm} f^{\lambda}(\omega) \operatorname{Tr}\left[\mathbf{G}^{\dagger}(t,\omega)\mathbf{G}(t,\omega) \mathbf{V}^{\lambda}(\omega) \right]\Biggr|_{0}^{T} \nonumber \\
	\therefore \langle \langle J_{11}^{\lambda}\rangle \rangle &=& 0\ .
\end{eqnarray}
Afterwards the second part~($\text{II}$) can be written as
\begin{eqnarray}
	\langle J_{12}^{\lambda}(t)\rangle &=& 2 e \int d\omega \hspace*{2 mm} f^{\lambda}(\omega) \int_{0}^{\infty} d \tau \operatorname{Re}\left[\operatorname{Tr}\left[\mathbf{G}^{\dagger}(t,\omega) e^{i \omega \tau} \mathbf{\Gamma}(\tau)\mathbf{G}(t-\tau,\omega) \mathbf{V}^{\lambda}(\omega) \right] \right], \nonumber \\
	\implies \langle \langle J_{12}^{\lambda} \rangle \rangle &=&\frac{1}{T} \int_{0}^{T} dt \hspace*{2 mm}\langle J_{12}^{\lambda}(t) \rangle \nonumber \\
	&=&  2 e \int d\omega \hspace*{2 mm} f^{\lambda}(\omega) \sum_{k}  \operatorname{Tr}\left[\mathbf{G}^{(k)\dagger}(\omega)  \left(\frac{\mathbf{\Gamma}(\omega + k\Omega)+\mathbf{\Gamma}^{\dagger}(\omega + k\Omega)}{2} \right)\mathbf{G}^{(k)}(\omega) \mathbf{V}^{\lambda}(\omega) \right]  \nonumber\\
	&=& -2\pi e \int d\omega \hspace*{2 mm}  \sum_{k,\lambda^{\prime}}  \operatorname{Tr}\left[\mathbf{G}^{(k)\dagger}(\omega)  \left(\mathbf{V}^{\lambda^{\prime}T} (\omega+k\Omega) \right)\mathbf{G}^{(k)}(\omega) \mathbf{V}^{\lambda}(\omega) \right]f^{\lambda}(\omega). \non
\end{eqnarray}
Thus adding the two contributions, we have
\begin{eqnarray}
	\langle \langle J_{1}^{\lambda} \rangle \rangle &=&\langle \langle J_{11}^{\lambda}  \rangle \rangle +\langle \langle J_{12}^{\lambda}  \rangle \rangle \non\\
	&=&-2\pi e \int d\omega \hspace*{2 mm}  \sum_{k,\lambda^{\prime}}  \operatorname{Tr}\left[\mathbf{G}^{(k)\dagger}(\omega)  \left(\mathbf{V}^{\lambda^{\prime}T} (\omega+k\Omega) \right)\mathbf{G}^{(k)}(\omega) \mathbf{V}^{\lambda}(\omega) \right]f^{\lambda}(\omega) \label{eq:j1}\ .
\end{eqnarray} 
Similarly, we can have~(from the second term of Eq.~(\ref{eq:raw_current_2}), $\la  J_{2}^{\lambda}(t) \ra$)
\begin{eqnarray}
	\langle \langle J_{2}^{\lambda} \rangle \rangle&=& 2\pi e \int d\omega \hspace*{2 mm}  \sum_{k,\lambda^{\prime}}  \operatorname{Tr}\left[\mathbf{G}^{(k)\dagger}(\omega)  \left(\mathbf{V}^{\lambda T} (\omega+k\Omega) \right)\mathbf{G}^{(k)}(\omega) \mathbf{V}^{\lambda^{\prime}}(\omega) \right]f^{\lambda^{\prime}}(\omega) \label{eq:j2}\ .
\end{eqnarray}
\end{widetext}
Therefore, the average current through the $\lambda$\textsuperscript{th} lead over a full time period is given by
\begin{eqnarray}
	\langle \langle J^{\lambda} \rangle \rangle&=&\langle \langle J_{1}^{\lambda} \rangle \rangle+\langle \langle J_{2}^{\lambda} \rangle \rangle \non\\
	&=& 2 \pi e \int d\omega \sum_{k} \left( \mathbf{T}_{\lambda \lambda^{\prime}}^{(k)} f^{\lambda^{\prime}}(\omega) - \mathbf{T}_{ \lambda^{\prime} \lambda}^{(k)} f^{\lambda}(\omega)\right), \non\\
	\label{eq:1}
\end{eqnarray}
where, the dynamical equivalence of transmission probability is defined as
\begin{equation}
	\mathbf{T}_{\lambda \lambda^{\prime}}^{(k)}(\omega)= \operatorname{Tr}\left[\mathbf{G}^{(k)\dagger}(\omega)  \left(\mathbf{V}^{\lambda T} (\omega+k\Omega) \right)\mathbf{G}^{(k)}(\omega) \mathbf{V}^{\lambda^{\prime}}(\omega) \right]. \label{eq:transmission}
\end{equation}
 Considering a two terminal setup~($\lambda=L$, and $\lambda^{\prime}=L,R$), we have total current through the left lead as
\begin{eqnarray}
	\langle \langle J^{L} \rangle \rangle \!\!
	&=& \!\! 2 \pi e \!\!\int \!\!d \omega \!\sum_{k} \! \left[\mathbf{T}_{LR}^{(k)} \hspace*{ 1 mm}f^{R}(\omega)\! -\mathbf{T}_{RL}^{(k)} f^{L}(\omega)\!\right] \nonumber \\
 &=&  \!\! 2 \pi e \!\!\int \!\!d \omega \!\sum_{k}  \mathbf{T}_{LL}^{(k)}\left[1- f^{L}(\omega)-f^{R}(-\omega)  \right]\non \\
 &+&\!\! 2 \pi e \!\!\int \!\!d \omega \!\sum_{k} \! \left[\mathbf{T}_{LR}^{(k)} \hspace*{ 1 mm}f^{R}(\omega)\! -\mathbf{T}_{RL}^{(k)} f^{L}(\omega)\!\right]
\label{eq:JL_full}.
\end{eqnarray} 
If the applied bias at the two leads $L$ and $R$ are $\pm V/2$, then the differential conductance at the left lead is given by
\begin{eqnarray}
	\sigma^{L}(V) &=& e \hspace*{ 2 mm} \frac{d \langle \langle J^{L} \rangle \rangle}{d \mu} \nonumber \\
 &=&  - 2 \pi e^{2} \!\!\int \!\!d \omega \!\sum_{k}  \mathbf{T}_{LL}^{(k)}\left[f^{\prime L}(\omega)+f^{\prime R}(-\omega)  \right]\non \\
 &+&\!\! 2 \pi e^{2} \!\!\int \!\!d \omega \!\sum_{k} \! \left[\mathbf{T}_{LR}^{(k)} \hspace*{ 1 mm}f^{\prime R}(\omega)\! -\mathbf{T}_{RL}^{(k)} f^{\prime L}(\omega)\!\right]\ .\non\\
 && \label{eq:sig_L}
\end{eqnarray}
In the flat band limit, lead density of states can be approximated as $\rho(\omega + k \Omega) \approx \rho=\text{constant}$ \ie $\mathbf{V}(\omega)\approx \mathbf{V}$. Using the above equation, 
and implementing non-equilibrium Green's function~(NEGF) technique, one can calculate differential conductance numerically. However, for one terminal setup (single lead contact), we can proceed 
further with an analytical approach. For single terminal setup, there is no right lead and the conductance becomes~\cite{KunduPRL2013}
\begin{eqnarray}
	\sigma(V) &=& -2 \pi e^{2}~ \mathbf{T}_{LL}^{(k)}~f^{\prime L}(\omega)\non\\
     &=&\!\!-2 \pi e^{2}\!\! \int d\omega \!\!\! \hspace*{2 mm}  \sum_{k}  \operatorname{Tr}\left[\mathbf{G}^{(k)\dagger}(\omega)  \mathbf{V}^{T}\mathbf{G}^{(k)}(\omega) \mathbf{V} \right] \! f^{'}(\omega)\non\\
	 \label{eq:mu_1_terminal}
\end{eqnarray}
If the system involves a superconductor, one has to add contributions arising from holes too. Then, the above expression~[Eq.~(\ref{eq:mu_1_terminal})] becomes 
\begin{eqnarray}
	\sigma(V) =  -2 \pi e^{2} \int d\omega \hspace*{2 mm}  \sum_{k}&&  \operatorname{Tr}\left[\mathbf{G}^{(k)\dagger}(\omega)  \mathbf{V}^{T}\mathbf{G}^{(k)}(\omega) \mathbf{V} \right] \non \\
	 && \times\left(f^{'}(\omega) +f^{'}(-\omega) \right)  \label{eq:mu_1_BDG}\ .
\end{eqnarray}
\begin{twocolumngrid}
Note that, $f(-\omega)$ in the above expression denotes Fermi-Dirac distribution for holes.
For one terminal setup the coupling matrix $\mathbf{K}$ vanishes except at the contact point i.e. $\mathbf{K}=\mathbf{K}^{c}\oplus 0 \oplus 0 \oplus.... N$ \text{times}, with $N$ being the system size. Consequently, $\mathbf{V}=\mathbf{V}^{c}\oplus 0 \oplus 0 \oplus.... N$ times.  Therefore. the coupling term at the contact site in electronic basis becomes $t_{h} a^{c \dagger} c^{c}$ (superscript $c \implies {\text {contact}}$) with $t_{h}$ being the hopping amplitude. For Rashba NW model~\cite{Mondal2023_NW} or Shiba chain model~\cite{Mondal_2023_Shiba}, one has to consider spin/chiral degrees of freedom in addition to electron-hole degrees of freedom. Thus the coupling matrix at contact site in BdG basis is given by
\begin{eqnarray}
\mathbf{K}^{c}=t_{h} ~\tau_{z} \otimes \sigma_{0}\ ,
\end{eqnarray} 
Then $\mathbf{V}^{c}$ is given by
\begin{eqnarray}
	\mathbf{V}^{c}&=& \mathbf{K}^{{c}\dagger} \rho \mathbf{K}^{c}\ , \nonumber \\
	&=& \rho t_{h}^{2} ~~\mathbf{I_{4}}  \label{eq:V}\ .
\end{eqnarray}
Let us define a new quantity: $\nu= 2 \pi \rho t_{h}^{2}$. Hence,  $\mathbf{V}^{c}$ can be written as
\begin{eqnarray}
	\mathbf{V}^{c}= \frac{\nu}{2 \pi} ~~ \mathbf{I_{4}}~=~\mathbf{V}^{cT} \label{eq:V_4by4}\ .
\end{eqnarray}
The structure of $\mathbf{V}$ renders convenient analytical handling of Eq.~(\ref{eq:mu_1_BDG}). As the non-vanishing contribution is arising from the contact site only, the non-trivial part 
of Eq.~(\ref{eq:mu_1_BDG}) is given by
\begin{eqnarray}
	\sigma(V)&=&\!\!-2 \pi e^{2}\!\! \int d\omega \!\!\! \hspace*{2 mm}  \sum_{k}  \operatorname{Tr}\left[\mathbf{G}^{c(k)\dagger}(\omega)  \mathbf{V}^{cT}\mathbf{G}^{c(k)}(\omega) \mathbf{V}^{c} \right] \non\\
	&& \times\left(f^{'}(\omega) +f^{'}(-\omega) \right)\ , \non  \\
	&=&\!\!-2 \pi e^{2}\!\!~ \left(\frac{\nu}{2 \pi}\right)^{2} \int d\omega \!\!\! \hspace*{2 mm}  \sum_{k}  \operatorname{Tr}\left[\mathbf{G}^{c(k)\dagger}(\omega) \mathbf{G}^{c(k)}(\omega) \right] \non\\
	&& \times\left(f^{'}(\omega) +f^{'}(-\omega) \right)\ ,
	\label{eq:mu_1_terminal_c}\
\end{eqnarray}
where, Green's function at contact site has the form~\cite{KunduPRL2013}
\begin{eqnarray}
\mathbf{G}^{c}=	\begin{pmatrix}
		G_{ee \ua \ua}^{c}&G_{ee \ua \da}^{c}&G_{eh \ua \ua}^{c}&G_{eh \ua \da}^{c}\\
		G_{ee \da \ua}^{c}&G_{ee \da \da}^{c}&G_{eh \da \ua}^{c}&G_{eh \da \da}^{c}\\
		G_{he \ua \ua}^{c}&G_{he \ua \da}^{c}&G_{hh \ua \ua}^{c}&G_{hh \ua \da}^{c}\\
		G_{he \da \ua}^{c}&G_{he \da \da}^{c}&G_{hh \da \ua}^{c}&G_{hh \da \da}^{c}\\
	\end{pmatrix}\ . \label{eq:Gc}
\end{eqnarray}
The trace operation in Eq.~(\ref{eq:mu_1_terminal_c}) for each spin component yields the expression
$|G_{ee}^{c}|^{2}+|G_{eh}^{c}|^{2}+|G_{he}^{c}|^{2}|G_{hh}^{c}|^{2}$. Using particle-hole symmetry, Eq.~(\ref{eq:mu_1_terminal_c}) becomes
\begin{widetext}
\begin{eqnarray}
	\sigma(V) &=& - \frac{e^{2} \nu^{2}}{2 \pi} \times2 \times \!\! \sum_{k,s,s^{\prime}} \int\!\!d\omega \!\! \hspace*{2 mm}  \left[ |G^{c(k)}_{ee,ss^{\prime}}(\omega)|^{2}+ |G^{c(k)}_{eh,ss^{\prime}}(\omega)|^{2}+ |G^{c(k)}_{he,ss^{\prime}}(\omega)|^{2} + |G^{c(k)}_{hh,ss^{\prime}}(\omega)|^{2} \right]  f^{'}(\omega)\ ,\non\\
	\label{eq:mu_2_BDG}
\end{eqnarray}	
where, $G^{c(k)}_{he,ss^{\prime}}$ corresponds to the hole~(with $s$-spin/chirality)-electron~(with $s^{\prime}$-spin/chirality) component  of Nambu-Gorkov Green's function for photon sector $k$ computed at the contact site. In the perturbative weak coupling limit~($\nu \ll t_{h}$), one can carry out perturbative analysis at the contact point to obtain the self-energy following~\cite{KunduPRL2013} as
\begin{eqnarray}
	\delta_{\bar{\alpha}} &=& - \langle \langle -\pi \left(\mathbf{V}+\mathbf{V}^{T} \right)  \rangle \rangle \nonumber 	 \\
	&=& \nu~\sum_{k}  \langle u_{\bar{\alpha}}^{c(k)} | \mathbf{I_{4}} | u_{\bar{\alpha}}^{c(k)} \rangle  \nonumber \\
	&=& \nu~ \sum_{k,s} \left[ |u_{\bar{\alpha},s}^{c(k)}|^{2}+ |v_{\bar{\alpha},s}^{c(k)}|^{2}\right]\ ,\label{eq:delta_alpha} 
\end{eqnarray}
where $u_{\bar{\alpha},s}^{c(k)}$, and $v_{\bar{\alpha},s}^{c(k)}$ are particle and hole components of the wave function, respectively, for spin/chiral sector $s$ and photon sector $k$ at the contact site. Therefore, in the zero temperature limit, one terminal differential conductance in presence of a bias voltage $V$ is given by
\begin{eqnarray}
	\tilde{\sigma}(V)&=& \operatorname{lim}_{V \rightarrow \epsilon_{\bar{\alpha}}} \frac{e^{2} \nu^{2}}{2 \pi}\times2  \sum_{k,s,s^{\prime}} \left[ |G^{c(k)}_{ee,ss^{\prime}}(V+n\Omega)|^{2}+|G^{c(k)}_{eh,ss^{\prime}}(V+n\Omega)|^{2}+|G^{c(k)}_{he,ss^{\prime}}(V+n\Omega)|^{2}+|G^{c(k)}_{hh,ss^{\prime}}(V+n\Omega)|^{2} \right]\nonumber \\
	&=&\frac{e^{2} \nu^{2}}{ \pi}  \!\!\sum_{n,k,s,s^{\prime},\bar{\alpha}} \left[\Biggr|\frac{\ket{u_{\bar{\alpha}}^{(n+k)}} \bra{u_{\bar{\alpha}}^{+(n)}}}{V-\epsilon_{\bar{\alpha}} +i \delta_{\bar{\alpha}}}\biggr|_{ee,ss^{\prime}}^{c} \Biggr|^{2}+\Biggr|\frac{\ket{u_{\bar{\alpha}}^{(n+k)}} \bra{u_{\bar{\alpha}}^{+(n)}}}{V-\epsilon_{\bar{\alpha}} +i \delta_{\bar{\alpha}}}\biggr|_{eh,ss^{\prime}}^{c} \Biggr|^{2}+\Biggr|\frac{\ket{u_{\bar{\alpha}}^{(n+k)}} \bra{u_{\bar{\alpha}}^{+(n)}}}{V-\epsilon_{\bar{\alpha}} +i \delta_{\bar{\alpha}}}\biggr|_{he,ss^{\prime}}^{c} \Biggr|^{2}+\Biggr|\frac{\ket{u_{\bar{\alpha}}^{(n+k)}} \bra{u_{\bar{\alpha}}^{+(n)}}}{V-\epsilon_{\bar{\alpha}} +i \delta_{\bar{\alpha}}}\biggr|_{hh,ss^{\prime}}^{c} \Biggr|^{2} \right] \nonumber \\
	&\approx& \frac{e^{2} \nu^{2}}{ \pi}  \sum_{n,k,s,s^{\prime},\bar{\alpha}} \left[ \Biggr| \frac{u_{\bar{\alpha},s}^{c (k+n)} u_{\bar{\alpha},s^{\prime}}^{c(n)}}{V-\epsilon_{\bar{\alpha}} +i \delta_{\bar{\alpha}}} \Biggr|^{2}+\Biggr| \frac{u_{\bar{\alpha},s}^{c (k+n)} v_{\bar{\alpha},s^{\prime}}^{c(n)}}{V-\epsilon_{\bar{\alpha}} +i \delta_{\bar{\alpha}}} \Biggr|^{2}+ \Biggr| \frac{v_{\bar{\alpha},s}^{c (k+n)} u_{\bar{\alpha},s^{\prime}}^{c(n)}}{V-\epsilon_{\bar{\alpha}} +i \delta_{\bar{\alpha}}} \Biggr|^{2}+\Biggr| \frac{v_{\bar{\alpha},s}^{c (k+n)} v_{\bar{\alpha},s^{\prime}}^{c(n)}}{V-\epsilon_{\bar{\alpha}} +i \delta_{\bar{\alpha}}} \Biggr|^{2} \right]\nonumber \\
	&=&\frac{e^{2} \nu^{2}}{ \pi}  \sum_{n,k,s,s^{\prime},\bar{\alpha}} \left[ \frac{ |u_{\bar{\alpha},s}^{c (k+n)} u_{\bar{\alpha},s^{\prime}}^{c(n)}|^{2}}{|V-\epsilon_{\bar{\alpha}} +i \delta_{\bar{\alpha}}|^{2}}+\frac{ |u_{\bar{\alpha},s}^{c (k+n)} v_{\bar{\alpha},s^{\prime}}^{c(n)}|^{2}}{|V-\epsilon_{\bar{\alpha}} +i \delta_{\bar{\alpha}}|^{2}}+\frac{ |v_{\bar{\alpha},s}^{c (k+n)} u_{\bar{\alpha},s^{\prime}}^{c(n)}|^{2}}{|V-\epsilon_{\bar{\alpha}} +i \delta_{\bar{\alpha}}|^{2}}+\frac{ |v_{\bar{\alpha},s}^{c (k+n)} v_{\bar{\alpha},s^{\prime}}^{c(n)}|^{2}}{|V-\epsilon_{\bar{\alpha}} +i \delta_{\bar{\alpha}}|^{2}} \right]\nonumber \\
	&=&
	\frac{2 e^{2}}{h} \times \sum_{n,\bar{\alpha},k,s,s^{\prime}} \frac{\nu^{2}}{\delta_{\bar{\alpha}}^{2}} \times \frac{ \left[ |u_{\bar{\alpha},s}^{c (k+n)} u_{\bar{\alpha},s^{\prime}}^{c(n)}|^{2}+
	|u_{\bar{\alpha},s}^{c (k+n)} v_{\bar{\alpha},s^{\prime}}^{c(n)}|^{2}+|v_{\bar{\alpha},s}^{c (k+n)} u_{\bar{\alpha},s^{\prime}}^{c(n)}|^{2}+|v_{\bar{\alpha},s}^{c (k+n)} v_{\bar{\alpha},s^{\prime}}^{c(n)}|^{2} \right]}{\left( \frac{V-\epsilon_{\bar{\alpha}}}{\delta_{\bar{\alpha}}} \right)^{2} + 1} \non \\
	&=&
	\sum_{n}\frac{2 e^{2}}{h} \!\!\! \sum_{\bar{\alpha},k,s,s^{\prime}} \! \frac{\nu^{2}}{\delta_{\bar{\alpha}}^{2}}\! \times\! \left[ |u_{\bar{\alpha},s}^{c (k+n)} u_{\bar{\alpha},s^{\prime}}^{c(n)}|^{2}+
	|u_{\bar{\alpha},s}^{c (k+n)} v_{\bar{\alpha},s^{\prime}}^{c(n)}|^{2}+|v_{\bar{\alpha},s}^{c (k+n)} u_{\bar{\alpha},s^{\prime}}^{c(n)}|^{2}+|v_{\bar{\alpha},s}^{c (k+n)} v_{\bar{\alpha},s^{\prime}}^{c(n)}|^{2} \right] \times \operatorname{L}\!\left( \frac{V-\epsilon_{\bar{\alpha}}}{\delta_{\bar{\alpha}}} \right) \non \\
	&=& \sum_{n} \sigma(V+n\Omega)\ , \label{eq:sum_slpit}
\end{eqnarray}
with $\sigma(V+n\Omega)=\frac{2 e^{2}}{h}  \sum_{\bar{\alpha},k,s,s^{\prime}} \frac{\nu^{2}}{\delta_{\bar{\alpha}}^{2}} \times \left[ |u_{\bar{\alpha},s}^{c (k+n)} u_{\bar{\alpha},s^{\prime}}^{c(n)}|^{2}+
|u_{\bar{\alpha},s}^{c (k+n)} v_{\bar{\alpha},s^{\prime}}^{c(n)}|^{2}+|v_{\bar{\alpha},s}^{c (k+n)} u_{\bar{\alpha},s^{\prime}}^{c(n)}|^{2}+|v_{\bar{\alpha},s}^{c (k+n)} v_{\bar{\alpha},s^{\prime}}^{c(n)}|^{2} \right] \times \operatorname{L}\left( \frac{V-\epsilon_{\bar{\alpha}}}{\delta_{\bar{\alpha}}} \right) \equiv \sigma^{(n)}(V)$. Here, the Lorentzian is defined as $\operatorname{L}(z)=\frac{1}{1+z^{2}}$.
\end{widetext}

\section{Extended~(frequency) space Hamiltonian}\label{sec:Extended space Hamiltonan}

In order to calculate the $\tilde{\sigma}$, we need the Floquet states in frequency space $|u_{\bar{\alpha}}^{(n)}\rangle$ and its components at contact sites. In this regard we diagonalize the extended space Hamiltonian following the Ref.~\cite{Eckardt_2015}. The extended space Hamiltonian has the form:
\begin{eqnarray}
	\!\!\!\begin{pmatrix}\!\!
		\ddots &\vdots & \vdots &\vdots & \vdots &\vdots & \!\!\!\!\iddots \\
		\hdots &H^{(0)}-2\Omega & H^{(-1)} & H^{(-2)} & H^{(-3)} & H^{(-4)} &\!\!\!\! \hdots\\
		\hdots \!\!\!&H^{(1)} &H^{(0)}-\Omega & H^{(-1)} & H^{(-2)} & H^{(-3)}  & \!\!\!\!\hdots\\
		\hdots &H^{(2)} &H^{(1)} &H^{(0)} & H^{(-1)} & H^{(-2)}  &\!\!\! \!\hdots\\
		\hdots  &H^{(3)} &H^{(2)} &H^{(1)} &H^{(0)}+ \Omega &H^{(-1)}   & \!\!\!\!\hdots\\
		\hdots \!\!\!\! &H^{(4)} &H^{(3)} &H^{(2)} &H^{(1)} &\!\!\!\! H^{(0)}+ 2\Omega   &\! \!\!\!\hdots\\
		\iddots &\vdots & \vdots &\vdots & \vdots &\vdots & \!\!\!\!\ddots \\
	\!\!\end{pmatrix}\ ,\non\\
 \!\!\!\!\!\label{eq:ext}
\end{eqnarray}
having block dimension $4N \!\!\!\! \times \!\!\!\! 4N$ with $H^{(n)}=\int_{0}^{T} \frac{dt}{T} e^{i n \Omega t} H(t)=H^{(-n)^{\dagger}}$. The form of these blocks depend on the nature of the external drive. We calculate the same for both of our driving protocols as mentioned for Rashba NW and helical Shiba chain models.

\subsection{Three step drive protocol}\label{sec:step}
For the step drive protocol introduced in Rashba NW case [Eq.~(\ref{eq:step_drive}) in the main text], the block $H^{(0)}$ in the diagonal is given by $\frac{1}{2}\left[H_{1}+H_{0}\right]$. 
However, off-diagonal blocks carry the form
\begin{eqnarray}
	H^{(n)}=\frac{(e^{\frac{i n \pi}{2}}-e^{\frac{i n 3\pi}{2}})}{i 2 \pi n} \left[H_{1}-H_{0}\right]\ . \label{eq:Hn_step}
\end{eqnarray}
Non-zero even values of $n$ yield $H^{(n=\rm{even})}=0$. However, for odd $n$, we have $H^{(\pm1)}=\frac{1}{\pi}\left[H_{1}-H_{0}\right]$, $H^{(\pm3)}=-\frac{1}{3\pi}\left[H_{1}-H_{0}\right]$, $H^{(\pm5)}=\frac{1}{5\pi}\left[H_{1}-H_{0}\right]$,.....etc. Note that, $H^{(n)}$ exhibits a $1/n$ fall.
\subsection{Sinusoidal drive protocol}\label{sec:sinusoidal}
In case of sinusoidal drive applied in Helical Shiba chain (magnet-superconductor hybrid structure)~[Eq.~(\ref{eq:sine_drive}) in the main text], the block in diagonal $H^{(0)}$ represents the static Hamiltonian $H_{0}$ itself. Also, only remaining non-vanishing off-diagonal blocks are $H^{(\pm 1)}$ which are given by $H^{(\pm 1)}=\frac{V_{0}}{2}\mathbf{I}_{N}\otimes \Gamma_{1}$.

Although the extended space Hamiltonian consists of infinite number of blocks, one can truncate the blocks for large $n$ as they do not effectively contribute to the emergent quasi-energies. For our numerical computations, we consider $n=-10$ to $n=10$ \ie total $2\times10+1=21$ photon sectors. Therefore, we effectively diagonalize $84N \times 84 N$~($21\times N \times 4 = 84 N$) matrix to obtain quasi-energies $\epsilon_{\bar{\alpha}}$ and quasi-states $|u_{\bar{\alpha}}^{(n)} \rangle$ where $N$ denotes the number of lattice sites in the given system. 

\end{twocolumngrid}
\bibliography{bibfile}{}

\end{document}